\documentclass{article}

\usepackage{arxiv}

\usepackage[utf8]{inputenc} 
\usepackage[T1]{fontenc}    
\usepackage{hyperref}       
\usepackage{url}            
\usepackage{booktabs,multirow}       
\usepackage{amsfonts,amsmath,amssymb,amsthm}       
\usepackage{nicefrac}       
\usepackage{microtype}      
\usepackage{cleveref}       
\usepackage{graphicx}
\usepackage{natbib}
\setcitestyle{numbers}
\usepackage{enumitem}
\usepackage[caption=false,font=normalsize,labelfont=sf,textfont=sf]{subfig}
\usepackage{algorithmic}

\numberwithin{equation}{section}
\theoremstyle{plain}

\theoremstyle{remark}

\newcommand{\bb}[1]{\boldsymbol{#1}}

\newcommand{\transpose}{^\mathsf{T}}

\DeclareMathOperator*{\argmin}{arg\,min}

\newcommand{\Hcal}{\mathcal{H}}

\title{{r}SVDdpd: A Robust Scalable Video Surveillance Background Modelling Algorithm}

\author{Subhrajyoty Roy\\
	Interdisciplinary Statistical Research Unit\\
	Indian Statistical Insititute, Kolkata \\
	\texttt{subhrajyoty\_r@isical.ac.in} \\
	\And
	Ayanendranath Basu\\
	Interdisciplinary Statistical Research Unit\\
	Indian Statistical Insititute, Kolkata \\
	\texttt{ayanbasu@isical.ac.in} \\
	\And
	Abhik Ghosh\\
	Interdisciplinary Statistical Research Unit\\
	Indian Statistical Insititute, Kolkata \\
	\texttt{abhik.ghosh@isical.ac.in} \\
}

\renewcommand{\shorttitle}{rSVDdpd: A Robust Scalable Video Surveillance Background Modelling Algorithm}

\hypersetup{
pdftitle={rSVDdpd: A Robust Scalable Video Surveillance Background Modelling Algorithm},
pdfauthor={Subhrajyoty Roy, Ayanendranath Basu, Abhik Ghosh},
pdfkeywords={Video Surveillance, Background Modelling, Object Tracking, Robust SVD},
}

\begin{document}
\maketitle

\begin{abstract}
    A basic algorithmic task in automated video surveillance is to separate background and foreground objects. Camera tampering, noisy videos, low frame rate, etc., pose difficulties in solving the problem. A general approach that classifies the tampered frames, and performs subsequent analysis on the remaining frames after discarding the tampered ones, results in loss of information. Several robust methods based on robust principal component analysis (PCA) have been introduced to solve this problem. To date, considerable effort has been expended to develop robust PCA via Principal Component Pursuit (PCP) methods with reduced computational cost and visually appealing foreground detection. However, the convex optimizations used in these algorithms do not scale well to real-world large datasets due to large matrix inversion steps. Also, an integral component of these foreground detection algorithms is singular value decomposition which is nonrobust. In this paper, we present a new video surveillance background modelling algorithm based on a new robust singular value decomposition technique rSVDdpd which takes care of both these issues. We also demonstrate the superiority of our proposed algorithm on a benchmark dataset and a new real-life video surveillance dataset in the presence of camera tampering. Software codes and additional illustrations are made available at the accompanying website rSVDdpd Homepage (https://subroy13.github.io/rsvddpd-home/)
\end{abstract}


\section{Introduction}\label{sec:introduction}

Automated surveillance from noisy videos is an extremely important problem with applications in different areas such as defence, security, research and monitoring, etc. In such video surveillance, the basic algorithmic task is to separate the background of a video from the foreground or moving objects, based on the input image frames from a surveillance video. The modelled background and foreground are then widely used in different image processing and computer vision applications. The most well-known and oldest applications were monitoring human activities in traffic surveillance systems. However, recently many other applications have been developed based on the same principle of background modelling such as detection of movements of animals and insects to study their behaviours, vision-based hand gesture recognition, autonomous vehicle pilot systems, content-based video coding, etc. Even though much work has been done towards solving this problem (see Garcia et al.~\cite{garcia2020background} for details), various challenges such as the presence of noisy frames, low frame rate of the camera, change in illumination, multimodal backgrounds, presence of small moving objects, camera tampering, etc., still present major hindrances. Mantini and Shah~\cite{mantini2019uhctd} define tampering as a change in the view of a surveillance camera, which may occur naturally or manually through unexpected or unauthorized actions. Reflection or glare of sunlight onto the camera lens, change in the camera view from daylight mode to night vision mode, defocusing of the camera lens, are some of the natural occurrences of tampering without any human intervention. On the other hand, intentionally covering or obstructing the view of the camera, rotating the camera to point in different directions, etc. are examples of manual tampering with malicious intent. The goal associated with the analysis of such video surveillance data in presence of camera tampering is primarily to classify the video frames as tampered or non-tampered ones; then the non-tampered frames are used for solving different computer vision problems of interest. Several approaches~\cite{mantini2019camera,sitara2019automated} have been developed to solve the above classification problem based on sophisticated image processing and deep learning techniques intended to detect the presence of such tampering. However, these techniques achieve good results only at the cost of an extensive amount of training data and computing power. Moreover, if the ultimate goal is to perform background modelling, discarding the tampered frames can result in loss of information, since if a frame is partially covered, the uncovered regions may still provide useful information including information about the unauthorized perpetrator. A robust methodology to extract the background and foreground should be able to make use of that information and provide valid and efficient inference under contaminated and noisy data.

Various methods for background modelling and foreground detection for video surveillance data have been developed in the past two decades. The different models can be classified into mathematical models (mean~\cite{lee2002background,mcivor2000background}, statistical~\cite{bouwmans2010statistical,lee2005effective}, fuzzy~\cite{subudhi2022kernel,zhao2012fuzzy}, Dempster-Schafer concepts~\cite{munteanu2015detection}), machine learning models (subspace learning~\cite{zhang2016riemannian,vaswani2018robust}, neural networks~\cite{maddalena2012sobs} and deep learning methods~\cite{giraldo2021graph,mandal2022empirical}), and signal processing methods (filter~\cite{Messelodi2005kalman}, sparse representation~\cite{stagiano2015online} and graph signal processing~\cite{giraldo2022graph}). Some of them are unsupervised methods as subspace learning methods, supervised as deep learning methods~\cite{mandal2022empirical} and semi-supervised as graph signal processing methods~\cite{giraldo2022graph}. The deep learning methods appear to offer the best performances; however, they require a large volume of labelled data and their performances decrease dramatically in the presence of unseen data. This is because these methods focus on improving the accuracy without modelling the underlying video data generation process in proper theoretical (or statistical) way. In contrast, unsupervised methods are backed by rich theoretical properties, require no labelled training data apriori and can be used in real-time applications. Among these, methods based on robust principal component analysis (PCA) have been shown to deliver state-of-the-art performances~\cite{bouwmans2014robust} compared to the traditional approaches of background modelling. These algorithms primarily rely on computing singular value decomposition (SVD) of some large matrices, which are highly susceptible to outliers. Usually, robust PCA algorithms circumvent this problem by computing truncated SVD multiple times and trimming any unreasonable solution to the factorization, due to which they become computationally expensive~\cite{candes2011,bouwmans2014robust,garcia2020background} and hence, of little practical use. Hence, while the deep learning methods are fast, scalable and parallelizable, they offer little theoretical justification and modelling of the background and foreground content. On the other hand, the unsupervised robust PCA based algorithms have a rich theoretical foundation, but they come at a high computational cost. In this paper, we propose a new background modelling and foreground estimation algorithm ``rSVDdpd'' based on a novel SVD methodology, which is unsupervised, theoretically justified, extremely robust and computationally efficient, thus correcting the deficiencies of each of the above approaches.

\subsection{Related Work}\label{sec:related-work}

Traditional approaches towards background modelling and foreground detection in video surveillance data followed various different approaches~\cite{bouwmans2014traditional}. Among classical techniques, the histogram based thresholding and running average methods~\cite{lee2002background,mcivor2000background} were computationally simple and fast, but the performances of such algorithms were not appealing enough for practical use. Over the course of time, several sophisticated foreground detection methods were introduced based on statistical models of the background image or video. A review and comparison of several useful statistical background modelling algorithms have been compiled by Bouwmans et al.~\cite{bouwmans2010statistical}. All of these models consider a decomposition of the video surveillance data 
\begin{equation}
    \bb{X} = \bb{L} + \bb{S}, 
    \label{eqn:base-decomposition}  
\end{equation}
\noindent where $\bb{X}$ is the data matrix obtained by stacking the pixel values in all frames of a video as columns. This means, for a $p$-frame long video with each frame being $h$ pixels long and $w$ pixels wide, the $\bb{X}$ matrix becomes of dimension $hw \times p$. The component $\bb{L}$ in Eq.~\eqref{eqn:base-decomposition} corresponds to the background content of the video, which is assumed to be generated from some statistical model. For instance, MOG~\cite{lee2005effective,caseiro2010foreground,caseiro2010background} and TensorMOG~\cite{ha2020tensormog} algorithms have been developed on the assumption that each column of $\bb{L}$ follows a mixture of Gaussian distribution. On the other hand, there exists another class of statistical background modelling algorithms derived from Principal Component Analysis (PCA) which models the matrix $\bb{L}$ as low rank. For instance, if the camera is static and is able to capture moderate to high resolution video feed, then it is expected that the background pixels from one frame to another frame will only differ in illumination changes. In this particular case, each column in the $\bb{L}$ matrix is expected to be proportional to each other, i.e., the matrix $\bb{L}$ is expected to be of rank one only. In general, since the subsequent frames in a video are correlated, a suitable low rank approximation $\bb{L}$ should be able to extract the non-moving part of the video, i.e., the background objects, while the difference ($\bb{X} - \bb{L}$) should be able to extract the sparse and noisy part, namely the moving foreground objects. While the classical SVD by itself can obtain such a decomposition of $\bb{X}$ and is computationally extremely fast, it cannot be directly used for foreground detection and background modelling. As noted by several authors~\cite{kumar2011new,niss2001,liu2003robust}, the SVD by itself is not a robust procedure and is prone to the different kinds of outliers such as shadows, illumination changes, camera tampering, jittering etc., which are common in real-life video surveillance data. Thus, the recent background modelling algorithms focus on obtaining a robust estimate of $\bb{L}$ through a robust PCA formulation. Such PCA based algorithms~\cite{zhang2016riemannian,babanezhad2018masaga} have received considerable attention over the last decade as they have been found to be better than MOG algorithms for several benchmark datasets~\cite{bouwmans2014robust} and also have broad applicability~\cite{xue2019side}. These algorithms minimize some objective function with respect to the eigenvalues and eigenvectors of $\bb{L}$, and the minimization is generally performed by a gradient descent algorithm on the Stiefel manifold (the Riemannian manifold of all orthogonal matrices of a fixed dimension) which is computationally extremely costly. Thus, although these methods produce superior results, they cannot be implemented in practice for real-time foreground detection of video surveillance data.

Based on the decomposition~\eqref{eqn:base-decomposition}, the ideal problem to solve for foreground detection would have been 
\begin{equation}
    \min_{\bb{L}, \bb{S}} \text{rank}(\bb{L}) + \lambda \Vert \bb{S}\Vert_0
    \label{eqn:ideal-problem}
\end{equation}
\noindent where $\lambda > 0$ is an arbitrary balance parameter. However, performing this minimization is NP-hard, hence a reasonable approximate solution would be to consider 
\begin{equation}
    \min_{\bb{L}, \bb{S}} \Vert \bb{L}\Vert_{\ast} + \lambda \Vert \bb{S}\Vert_1
    \label{eqn:PCP-problem}
\end{equation}
\noindent where $\Vert \bb{L}\Vert_{\ast}$ denotes the nuclear norm (i.e., the sum of singular values) of $\bb{L}$. Candes et al.~\cite{candes2011} define this problem as ``Principal Component Pursuit'' (PCP) and under the assumption that $\bb{L}$ is exactly low rank and $\bb{S}$ is exactly sparse, they are able to show that the solution of the approximate problem~\eqref{eqn:PCP-problem} perfectly recovers a solution of the ideal problem~\eqref{eqn:ideal-problem}. Since this method, too, was computationally costly, numerous algorithms have been built to solve the Principal Component Pursuit problem as in Eq.~\eqref{eqn:PCP-problem} with less computational cost. Some optimization techniques like Augmented Lagrangian method~\cite{shen2014augmented}, Gradient descent method~\cite{yi2016fast}, etc. have proven useful in this regard. Xu et al.~\cite{xu2012robust} considered another variant of the ideal problem~\eqref{eqn:ideal-problem} called Outlier Pursuit (OP)
\begin{equation}
    \min_{\bb{L}, \bb{S}} \Vert \bb{L}\Vert_{\ast} + \lambda \Vert \bb{S}\Vert_{1, 2}
    \label{eqn:OP-problem}
\end{equation}
\noindent where $\Vert \bb{S}\Vert_{1, 2}$ is the sum of the $L_2$ norms of the columns of $\bb{S}$. One limitation of PCP and OP is that they assume the decomposition of the data $\bb{X}$ into the low rank matrix and the sparse matrix to be exact, but real life videos hardly satisfy that due to quantization of pixel values, measurement errors, shadows, artifacts, slowly moving backgrounds, etc. Therefore, a more realistic decomposition of the video surveillance data $\bb{X}$ is to consider 
\begin{equation}
    \bb{X} = \bb{L} + \bb{S} + \bb{N}
    \label{eqn:error-decomposition}
\end{equation}
\noindent where $\bb{N}$ is a small perturbation matrix with independent elements, which is neither low rank nor sparse. Babacan et al.~\cite{babacan2012sparse} apply a Bayesian modelling to the decomposition~\eqref{eqn:error-decomposition} and use Gibbs sampling to obtain posterior estimates of the low rank matrix $\bb{L}$ through sampling in the Stiefel manifold. In contrast, Liu et al.~\cite{liu2017sparsity} used the Augmented Lagrangian method repeatedly to solve the problem 
\begin{equation}
    \min_{\bb{L}, \bb{S}} \Vert \bb{L}\Vert_{\ast} + \beta \Vert \bb{S} \Vert_0 + \lambda \Vert \bb{X} - \bb{L} - \bb{S}\Vert_{1}
\end{equation} 
\noindent where both $\lambda, \beta > 0$ are penalty parameters that can be tuned.

\subsection{Issues with existing works}

Despite being extremely powerful, the robust PCA algorithms by themselves have several bottlenecks which limit their practical use. Various modifications of these algorithms have been proposed in the last decade including incremental algorithms~\cite{narayan2018merop,he2012incremental,rodriguez2016incremental} to identify foreground content as soon as new video frames are recorded, spatio-temporal algorithms~\cite{javed2018moving,zhu2020motion} which enable foreground detection to be motion aware, and algorithmic improvements~\cite{cai2019accelerated,xiu2020lrpca} to address scalability issues so that these algorithms can be used to estimate the background content in a large scale high-resolution video surveillance data within a reasonable time. There are two primary bottlenecks in reducing the computational complexity of these algorithms. One, these algorithms rely on the convex optimization problems as in Eq.~\eqref{eqn:PCP-problem} or \eqref{eqn:OP-problem} which present significant challenges in optimization in a high dimensional setting~\cite{jain2017nonconvex}. Two, these algorithms use numerous truncated SVD, matrix multiplication and matrix inversion steps which have currently a computational complexity of $O(n^{2.37188})$~\cite{duan2022faster} for $n \times n$ matrices, and the order can grow rapidly and significantly for large values of $n$ in high dimensional matrices. To address both of these issues, we propose a novel algorithm to compute robust estimates of the singular values and vectors themselves, without requiring the formulation of a convex optimization problem as PCP, but rather a statistical robust regression problem. We also present an iterative algorithm based on the alternating regression trick, which circumvents the aforementioned computationally intensive steps and reduces the computational complexity to only $O(n^2)$ in each iteration for an $n \times n$ matrix. Also, the iteration steps of the proposed algorithm are highly parallelizable, hence the algorithm can scale efficiently to arbitrarily high dimensional video surveillance data, within a reasonable amount of time, provided necessary memory requirements.

For example, as presented in Candes et al.~\cite{candes2011}, it takes about $13$ seconds per frame for the exact PCP method to converge for a $176\times 144$ low-resolution grayscale video, and clearly, this high computational complexity hinders its practical use despite having good performance for foreground extraction. In comparison, GRASTA algorithm~\cite{he2012incremental} which does not take the robust PCA approach takes only $0.13$ seconds per frame to extract the foreground content of a video of the same resolution, but its performance accuracy does not match that of the robust PCA via the PCP approach~\cite{bouwmans2014robust}. On the other hand, our proposed algorithm takes only $0.182$ seconds per frame to separate the background and foreground of a video of the same dimensionality and achieves the same level of accuracy as the exact PCP method.

\subsection{Our contributions}

\begin{figure*}[!t]
    \centering
    \subfloat[]{
        \includegraphics[width=0.19\linewidth]{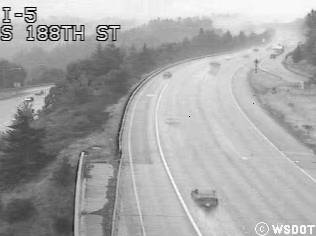}
        \includegraphics[width=0.19\linewidth]{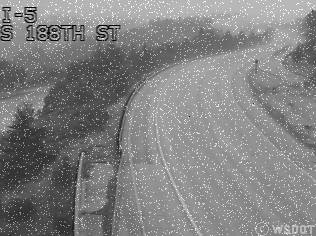}
        \includegraphics[width=0.19\linewidth]{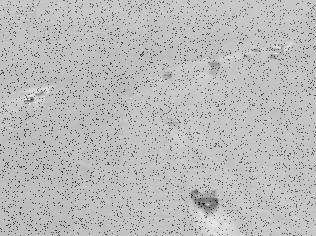}
        \includegraphics[width=0.19\linewidth]{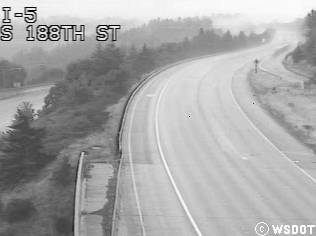}
        \includegraphics[width=0.19\linewidth]{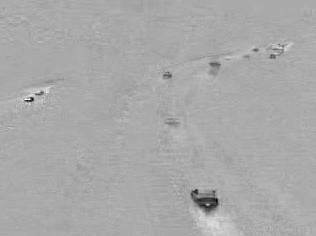}
    }
    \hfill
    \subfloat[]{
        \includegraphics[width=0.19\linewidth]{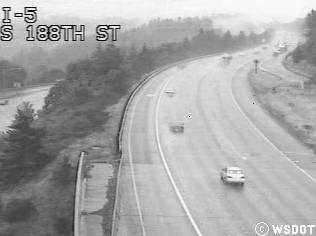}
        \includegraphics[width=0.19\linewidth]{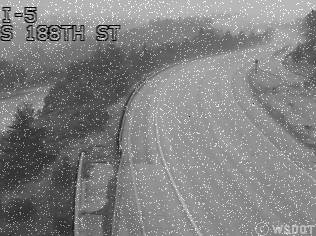}
        \includegraphics[width=0.19\linewidth]{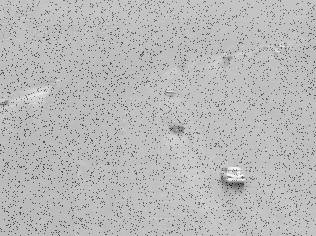}
        \includegraphics[width=0.19\linewidth]{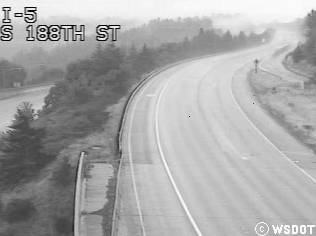}
        \includegraphics[width=0.19\linewidth]{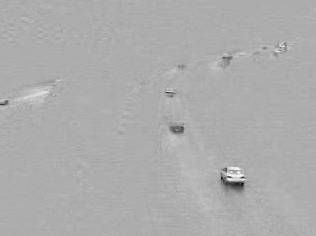}
    }
    \hfill
    \subfloat[]{
        \includegraphics[width=0.19\linewidth]{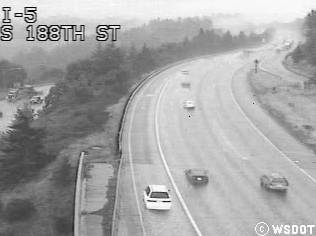}
        \includegraphics[width=0.19\linewidth]{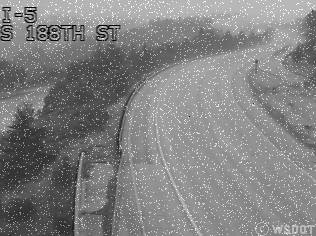}
        \includegraphics[width=0.19\linewidth]{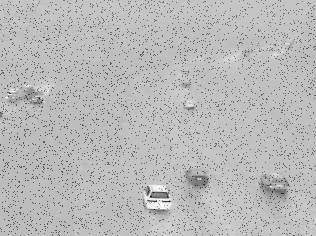}
        \includegraphics[width=0.19\linewidth]{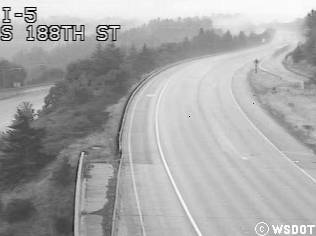}
        \includegraphics[width=0.19\linewidth]{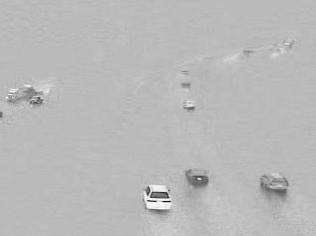}
    }
    \caption{The true images, background and foreground images estimated by usual SVD and the proposed rSVDdpd method (from left to right in a row), corresponding to frame 5 (a), 18 (b) and 42 (c) of the ``freeway'' video sequence. Enlarged better quality images are available online at https://subroy13.github.io/rsvddpd-home/.}
    \label{fig:video-freeway}
\end{figure*}

In this paper, we aim to provide a novel background modelling and foreground extraction algorithm ``rSVDdpd'' via an alternative robust singular value decomposition methodology based on the popular density power divergence. The minimum density power divergence introduced by Basu et al.~\cite{basu-dpd} has been proven to be robust and efficient in estimation in several contexts~\cite{ghosh2013,xiong2021minimum}. We leverage its robustness properties by directly applying it to the video surveillance background modelling problem, where the noisy and the sparse content can be viewed as outliers present in the data. As an illustration, we consider the ``freeway'' video sequence from UCSD Background Subtraction Dataset~\cite{ucsd-background-data} consisting of $44$ frames of grayscale images of dimension $316 \times 236$. It depicts a highway in the background with moving cars as foreground objects having different shades of gray. One can use the usual singular value decomposition (SVD) to extract such background and foreground content, but that usually becomes highly susceptible to noise present in the video surveillance data. Here, we add salt-and-pepper noise to only a few of the consecutive frames of the video as mild tampering, which ideally should affect only the tampered frames but not to the entire video. As shown in Figure~\ref{fig:video-freeway}, the usual SVD based background model works poorly even in non-tampered frames. In contrast, the proposed robust algorithm rSVDdpd is able to extract the foreground and background content very efficiently, by removing noisy artifacts such as moving shadows, illumination change, etc. Also, rSVDdpd solves a non-convex optimization problem by an alternating fixed point iteration method. Each step of the iteration is computationally extremely simple to perform, requiring only scalar inner product computations, which are highly parallelizable making it suitable for very high dimensional or high resolution video surveillance applications.

In summary, we aim to make the following contributions in this paper.
\begin{itemize}
    \item Propose a fast, robust and scalable alternative to the video surveillance background modelling problem through robust singular value decomposition.
    \item We compare the proposed rSVDdpd algorithm with several existing background modelling algorithms on the Background Model Challenge Benchmark dataset to show its effectiveness.
    \item We also demonstrate how rSVDdpd can obtain reliable background and foreground estimates in the presence of camera tampering by showing its performance on a novel real-life ``University of Houston Camera Tampering Dataset'' (UHCTD).
    \item In this paper, we also discuss various desirable theoretical properties the rSVDdpd estimate enjoys.
\end{itemize}

The rest of the paper is organized as follows. In Section~\ref{sec:proposed-method}, we develop our background modelling algorithm from a regression based and a penalty based approach. Following this development, we present the convergence results and some basic mathematical properties of the proposed rSVDdpd estimator in Section~\ref{sec:math-property}. Section~\ref{sec:experiments} contains the experimental results regarding the performance comparison of different background modelling algorithms on the Background Model Challenge Benchmark dataset. Finally, in Section~\ref{sec:real-dataset}, we also make a similar comparison on a novel real-life dataset UHCTD to demonstrate the superiority of our method in the presence of camera tampering, and some concluding remarks are presented in Section~\ref{sec:conclusion}.

\section{Proposed rSVDdpd Method}\label{sec:proposed-method}

Instead of the decomposition as in Eq.~\eqref{eqn:error-decomposition}, we combine $\bb{S}$ and $\bb{N}$ into a single error matrix $\bb{E}$, whose elements are small independent noise component except certain spikes due to the sparse component $\bb{S}$. Let the video surveillance data matrix $\bb{X}$ be of dimension $n \times p$ ($n$ and $p$ may be different) and the low rank component $\bb{L}$ is of rank $r$, where $r$ is much lower than $n$ or $p$. Then, the decomposition of $\bb{X}$ can be expressed in terms of the singular values and vectors of $\bb{L}$, namely
\begin{equation}
    \bb{X} = \sum_{k = 1}^{r} \lambda_k \bb{u}_{k}\bb{v}_{k}\transpose + \bb{E},    
\end{equation}
\noindent where $\bb{u}_{k}$ is a vector of length $n$ and $\bb{v}_{k}$ is a vector of length $p$ for $k = 1, 2, \dots r$, and $\bb{E}$ is a $n\times p$ matrix. The entries of the error matrix $\bb{E}$, i.e., $e_{ij}$s are generally expected to be smaller in magnitude than the corresponding entry of the data matrix $\bb{X}$, except for some outlying values corresponding to the pixels of the foreground object in different frames. The left singular vectors $\{\bb{u}_k\}_{k = 1, 2, \dots r}$'s and the right singular vectors $\{\bb{v}_k\}_{k = 1, 2, \dots r}$'s also satisfy the orthonormality constraints, i.e.,
\begin{align*}
    \bb{u}_{k}\transpose \bb{u}_{l} 
     = \begin{cases}
        1 & \text{ if } k = l\\
        0 & \text{ if } k \neq l
    \end{cases}, 
    \quad
    \bb{v}_{k}\transpose \bb{v}_{l} 
    & = \begin{cases}
        1 & \text{ if } k = l\\
        0 & \text{ if } k \neq l
        \end{cases}\\
    & \qquad k, l = 1, 2, \dots r,
\end{align*}
\noindent and the singular values $\lambda_k$ satisfy the nonnegativity constraints. As in~\cite{rey2007-total-svd}, the above description can also be reformulated into a matrix factorization as
\begin{equation}
  \bb{X} = \sum_{k = 1}^{r} \bb{a}_{k}\bb{b}_{k}\transpose + E,
  \label{eqn:svd}    
\end{equation}
\noindent where $\bb{a}_k$s and $\bb{b}_k$s are still orthogonal sets of vectors for $k = 1, 2, \dots r$, but not necessarily normalized. Once the estimates of $\bb{a}_k$s and $\bb{b}_k$s are known, they can be normalized to obtain the $\bb{u}_k$s and $\bb{v}_k$s and the singular values are then given by $\lambda_k = \Vert \bb{a}_k \Vert \Vert \bb{b}_k \Vert$ for each $k=1, \ldots, r$, where $\Vert\cdot\Vert$ denotes the usual Euclidean ($L_2$) norm. 

For the computation of the left and right singular vectors from Eq.~\eqref{eqn:svd}, one can proceed sequentially focusing only on a rank one approximation as follows. First, consider the approximation $\bb{X} \approx \bb{a}_1\bb{b}_1\transpose$. Once we have estimates for $\widehat{\bb{a}}_1$ and $\widehat{\bb{b}}_1$, we can obtain $\widehat{\lambda}_1 = \Vert \widehat{\bb{a}}_1 \Vert \Vert \widehat{\bb{b}}_1\Vert$, and normalize $\widehat{\bb{a}}_1$ and $\widehat{\bb{b}}_1$ as required. Next, we compute the residual matrix $\bb{X} - \widehat{\bb{a}}_1\widehat{\bb{b}}_1\transpose$, and the rank one approximation algorithm can be used again on this residual matrix, to obtain the second singular value and the corresponding vectors. Proceeding similarly, one can compute all $r$ singular values and singular vectors of the given data matrix $\bb{X}$. Therefore, for ease of explanation, we will describe the algorithm only for obtaining the vectors $\bb{a}$ and $\bb{b}$ yielding the best rank-one approximation of the data matrix $\bb{X}$ as
\begin{equation}
    \bb{X} \approx \bb{a} \bb{b}\transpose.
    \label{eqn:svd-rank-one}
\end{equation}
\noindent Accordingly, in the following, we shall use $a_i$ for $i = 1, 2, \dots n$, to denote the elements of the vector $\bb{a}$ and similarly, $b_j$ for $j = 1, 2, \dots p$, to denote the elements of the vector $\bb{b}$. The elements of the $\bb{X}$ matrix are denoted by $x_{ij}$.

\subsection{The Regression based approach}\label{sec:regression-approach}

We consider the idea of transforming the SVD estimation problem into a regression problem, following~\cite{rey2007-total-svd}. For this purpose, let us first fix the index $j$ (i.e., the column index), and consider the setup
\begin{equation}
    X_{ij} = a_i b_j + e_{ij}, \qquad i = 1, 2, \dots n,
    \label{eqn:regression-reduced}
\end{equation}
\noindent which is a simple linear regression problem without any intercept. Here, $x_{ij}$s are observed response variables and $e_{ij}$s are random error components. Therefore, for any given value of $\bb{a}$, we can treat $a_i$s as covariate values and estimate the $b_j$s as regression slopes, and as we vary the column index $j = 1, 2, \dots p$, we are posed with $p$ such linear regression problems, jointly yielding an estimate of $\bb{b}=(b_1, \ldots, b_p)$. Next, given these estimated values of $b_j$s, we treat them as covariates in Eq.~\eqref{eqn:regression-reduced} and estimate the $a_i$s as the unknown regression parameters for each $i=1, \ldots, n$. Repeating these two steps sequentially until convergence, we get the final desired estimates of $\bb{a}$ and $\bb{b}$ from which the desired singular value and vectors can be obtained as discussed previously.

Now, for estimating the regression coefficients ($b_j$ or $a_i$) in each iteration of the aforementioned alternating regression approach, we propose to use the robust and efficient minimum Density Power Divergence (DPD) estimation procedure. The minimum DPD estimator (MDPDE) was initially developed for independent and identically distributed data in~\cite{basu-dpd}, and later extended to the more general independent non-homogeneous set-up in~\cite{ghosh2013}. It considers a general form of DPD-based loss function involving 
\begin{equation}
    V_i(y_i; \bb{\theta}) = \left[ \int f_{i}(y; \bb{\theta})^{1 + \alpha} dy - \left( 1 + \frac{1}{\alpha} \right) f_{i}(y_{i}; \bb{\theta})^{\alpha} \right],
    \label{eqn:v-theta-general}
\end{equation}
\noindent where $\alpha > 0$ is the robustness tuning parameter, and the distribution of the $i$-th sample observation $y_i$ is modeled using a parametric family of distributions $\mathcal{F}_i = \{f_i(y; \bb{\theta}) : \bb{\theta} \in \bb{\Theta} \}$ for each $i$. As a particular case, the MDPDE under the linear regression model with normally distributed errors are also discussed in~\cite{ghosh2013}, which can be used in the present context. For our purpose, as in the classical setup of the robust statistics, we can model the error components $e_{ij}$s in Eq.~\eqref{eqn:regression-reduced} to be independent and normally distributed with mean zero and variance $\sigma^2$, except for some outliers due to the spikes that are present in the $\bb{E}$ matrix. Therefore, the MDPDE objective function for estimating the unknown parameters $a_i$ or $b_j$ and $\sigma^2$ from Eq.~\eqref{eqn:regression-reduced}, turns out to be 
\begin{equation}
    H_{n, p}(\bb{a}, \bb{b}, \sigma^2) = \dfrac{1}{np} \sum_{i = 1}^n \sum_{j = 1}^p V(x_{ij}; a_i, b_j, \sigma^2),
    \label{eqn:H-theta}
\end{equation}
\noindent where 
\begin{multline}
    V(y; c, d, \sigma^2) = (2\pi)^{-\alpha/2} \sigma^{-\alpha} \left[ (1+\alpha)^{-1/2} \right.\\ 
    \left. - \frac{(1 + \alpha)}{\alpha} \exp\left\{ -\alpha \frac{(y - cd)^2}{2\sigma^2} \right\} \right],
    \label{eqn:v-theta-normal}
\end{multline}
\noindent is a simplification of Eq.~\eqref{eqn:v-theta-general} with $f_i(y, \bb{\theta})$ as the density of a normal distribution with mean $cd$ and variance $\sigma^2$. The MDPDE of the parameter $\bb{\theta} = (\bb{a}, \bb{b}, \sigma^2)$ is then obtained by minimizing this objective function in Eq.~\eqref{eqn:H-theta} with respect to each of the components of the triplet. So, instead of estimating the elements of the scaled singular vectors $\bb{a}$ and $\bb{b}$, i.e., $a_i$s and $b_j$s in the alternating regression models, we can indeed minimize this MDPDE objective function iteratively over either of the parameters ($a_i$, $b_j$ or $\sigma^2$) given the most recent estimates of the other parameters. By standard differentiation, as in~\cite{ghosh2013}, the corresponding estimating equations turn out to be
\begin{align}
    \sum_{j = 1}^{p} b_j (x_{ij} - a_i b_j) e^{-\alpha \frac{(x_{ij} - a_ib_j)^2}{2\sigma^2}} & = 0, \ i = 1, \ldots n, \label{eqn:estimating-eqn-1}\\
    \sum_{i = 1}^{n} a_i (x_{ij} - a_i b_j) e^{-\alpha \frac{(x_{ij} - a_ib_j)^2}{2\sigma^2}} & = 0, \ j = 1, \ldots p, \label{eqn:estimating-eqn-2}\\
    \sum_{i = 1}^{n}\sum_{j = 1}^{p} \left[ 1 -  \frac{(x_{ij} - a_ib_j)^2}{\sigma^2} \right] e^{-\alpha \frac{(x_{ij} - a_ib_j)^2}{2\sigma^2}} & = \frac{\alpha}{(1 + \alpha)^{3/2}}.
    \label{eqn:estimating-eqn-3}
\end{align}
From another viewpoint, the mentioned proposal yields a robust generalization of the ususal SVD estimators for $\alpha > 0$. To see this, denoting $w_{ij} = e^{- \alpha \frac{(x_{ij} - a_ib_j)^2}{2\sigma^2}}$, it follows that the estimating equations in Eq.~\eqref{eqn:estimating-eqn-1}-\eqref{eqn:estimating-eqn-3} can be rearranged into the form
\begin{equation}
    \begin{split}
        a_i & = \dfrac{\sum_j b_j x_{ij} w_{ij} }{ \sum_j b_j^2 w_{ij} }, \ i = 1, \ldots n; \\
        b_j & = \dfrac{\sum_i a_i x_{ij} w_{ij} }{ \sum_i a_i^2 w_{ij} }, \ j = 1, \ldots p; \\
        \sigma^2 & = \dfrac{\sum_i \sum_j (x_{ij} - a_i b_j)^2 w_{ij}}{ \sum_i \sum_j w_{ij} - \frac{\alpha}{(1 + \alpha)^{3/2}} }.        
    \end{split}
    \label{eqn:algo-eqn}
\end{equation}
\noindent In the first estimating equation of Eq.~\eqref{eqn:algo-eqn}, as $x_{ij} \approx a_i b_j$, i.e., $x_{ij}/b_j \approx a_i$, the equation considers the weighted mean of $x_{ij}/b_j$ with weights $b_j^2w_{ij}$. Similar normalization is also present in the second equation. In the third equation, we see that $\sigma^2$ is approximately a normalized version of the squared residuals $(x_{ij} - a_ib_j)^2$ except for a term subtracted from the denominator to make the corresponding estimating equation unbiased. Thus, each of the parameters are estimated as a weighted average of different crude estimates based on the elements of the data matrix $\bb{X}$ and the most recent values of the other parameters. Also, for any $\alpha > 0$, $w_{ij} = e^{- \alpha\frac{(x_{ij} - a_ib_j)^2}{2\sigma^2}}$ is a decreasing function of the magnitude of the residuals, i.e., of $\vert x_{ij} - a_ib_j \vert$. This decreasing nature is crucial to make the overall estimates robust since a large deviation from the model having large $\vert x_{ij} - a_i b_j \vert$ would yield smaller weights, and hence those outlying values would have little effect on the estimating equations given in Eq.~\eqref{eqn:algo-eqn}.

The particular form of these weights $w_{ij}$ in Eq.~\eqref{eqn:algo-eqn} appears due to the assumption of normality of the errors $e_{ij}$. In a more general version with appropriately modified weight function, one can consider the estimating equations
\begin{align}
    a_i & = \dfrac{\sum_j b_j x_{ij} \psi(\vert x_{ij} - a_ib_j \vert ) }{ \sum_j b_j^2 \psi(\vert x_{ij} - a_ib_j \vert ) }, \qquad i = 1, \ldots n, 
    \label{eqn:general-1}\\
    b_j & = \dfrac{\sum_i a_i x_{ij} \psi(\vert x_{ij} - a_ib_j \vert ) }{ \sum_i a_i^2 \psi(\vert x_{ij} - a_ib_j \vert ) }, \qquad j = 1, \ldots p,
    \label{eqn:general-2}\\
    \sigma^2 & = \dfrac{\sum_i \sum_j (x_{ij} - a_i b_j)^2 \psi(\vert x_{ij} - a_ib_j \vert )}{ \sum_i \sum_j \psi(\vert x_{ij} - a_ib_j \vert ) - S_\psi }.
    \label{eqn:general-3}
\end{align}
\noindent where $\psi(\cdot)$ is a suitably smooth and decreasing function of its argument, and $S_\psi$ is a suitably chosen quantity to make the estimating equation corresponding to $\sigma^2$ unbiased.

\subsection{Orthogonalization of Singular Vectors}\label{sec:penalized-regression-approach}

In the mathematical framework developed in Section~\ref{sec:regression-approach}, there is no component in the regression formulation that ensures orthogonality of the singular vectors. It is natural to extend the objective function $H_{n,p}(\bb{a}, \bb{b}, \sigma^2)$ of Eq.~\eqref{eqn:H-theta} using a penalty function to achieve the orthogonality requirements. One such penalty function could be the sum of the squared inner products (dot-products) of the current singular vector with all preceding singular vectors. 

To formulate this penalized approach, let us assume, $\bb{a}_1, \bb{a}_2, \dots , \bb{a}_k$ be the $k$ non-normalized left singular vectors and $\bb{b}_1, \bb{b}_2, \dots, \bb{b}_k$ be the $k$ non-normalized right singular vectors of the data matrix $\bb{X}$, leading to the singular values $\lambda_l = \Vert \bb{a}_l \Vert \Vert \bb{b}_l \Vert$ for $l = 1, \ldots, k$. This amounts to assuming that we already have the rank $k$ approximation of the data matrix as
\begin{equation*}
    \bb{X} \approx \sum_{r=1}^{k} \bb{a}_r \bb{b}_r\transpose.
\end{equation*}
\noindent With these already estimated (non-normalized) singular vectors $\bb{a}_1, \dots \bb{a}_k, \bb{b}_1, \dots \bb{b}_k$, the task of obtaining the next orthogonal (non-normalized) singular vectors $\bb{a}_{(k+1)}$ and $\bb{b}_{(k+1)}$ (assume $k \leq \text{rank}(\bb{X}) - 1$) becomes the same as finding a rank one decomposition like Eq.~\eqref{eqn:svd-rank-one} of the transformed data matrix $\bb{X}' = \left( \bb{X} - \sum_{r = 1}^k \bb{a}_r\bb{b}_r\transpose \right)$ subject to the conditions
\begin{equation*}
    \bb{a}\transpose \bb{a}_r = \bb{b}\transpose \bb{b}_r = 0, \qquad r = 1, 2, \dots k.
\end{equation*}
\noindent Thus, we can again use the alternating regression approach to estimate the parameters $\bb{a}, \bb{b}, \sigma^2$ iteratively and consider the penalized MDPDE objective function with Lagrangian parameter $\xi$ as
\begin{multline}
    \widetilde{H}_{k; n, p}(\bb{a}, \bb{b}, \sigma^2) = \dfrac{1}{np} \sum_{i = 1}^n \sum_{j = 1}^p V\left(x'_{ij}; a_i, b_j, \sigma^2\right)\\
    + \dfrac{\xi}{k} \left[ \sum_{r = 1}^k \dfrac{(\bb{a}\transpose \bb{a}_r)^2}{\Vert \bb{a}_r \Vert^2} + \sum_{r = 1}^k \dfrac{(\bb{b}\transpose \bb{b}_r)^2}{\Vert \bb{b}_r \Vert^2} \right],\\
    k = 1, 2, \dots \text{rank}(\bb{X}-1),
    \label{eqn:H-theta-penalized}
\end{multline}
\noindent where $x'_{ij}$ is the $(i,j)$-th element of $\bb{X}'$. Notice that the normalization factors $\Vert \bb{a}_r\Vert^2$ and $\Vert \bb{b}_r \Vert^2$ have been used in Eq.~\eqref{eqn:H-theta-penalized} to make sure that each penalty term is of similar magnitude.

Again, differentiating the objective function in Eq.~\eqref{eqn:H-theta-penalized} with respect to the individual parameters and setting them equal to 0, yields the estimating equations for the parameters $\bb{a}, \bb{b}, \sigma^2$. A reorganization of the estimating equations lead to a fixed point iteration formula similar to Eq.~\eqref{eqn:algo-eqn} as given by
\begin{equation}
    \begin{split}
        a_i
        & = \left( \sum_{j = 1}^p b_j^2 w_{ij}\right)^{-1} \left[ \sum_{j = 1}^p b_j x'_{ij} w_{ij} + C_{\xi} \sum_{r=1}^{k} (\bb{a}_r\transpose \bb{a}) (\bb{a}_r)_i \right], \\ 
        & \qquad \qquad  i = 1, \dots n,\\
        b_j 
        & = \left( \sum_{i = 1}^n a_i^2 w_{ij}\right)^{-1} \left[ \sum_{i = 1}^n a_i x'_{ij} w_{ij} + C_{\xi} \sum_{r=1}^{k} (\bb{b}_r\transpose \bb{b}) (\bb{b}_r)_j \right],\\ 
        &  \qquad \qquad  j = 1, \dots p,\\
        &  \qquad \qquad k = 1, 2, \dots \text{rank}(\bb{X})-1,
    \end{split}
    \label{eqn:est-eqn-penalty}
\end{equation}
\noindent where $w_{ij} = e^{ -\alpha \frac{(x'_{ij} - a_i b_j)^2}{2\sigma^2}}$, and $C_{\xi} = \frac{np(2\pi)^{(\alpha/2)} \sigma^\alpha \xi}{k(1 + \alpha)}$. Since, the penalty term is not a function of $\sigma^2$, the estimating equation corresponding to $\sigma^2$ remains same as in Eq.~\eqref{eqn:algo-eqn}.

Unfortunately, the convergence of the algorithm with estimating equations Eq.~\eqref{eqn:est-eqn-penalty} is highly sensitive to the choice of Lagrangian penalty parameter $\xi$. Often, the choice of $\xi$ can be obtained through a cross validation approach as in~\cite{owen2009bi}. However, this adds another level of complexity to the algorithm making it computationally expensive. Since, our aim is to obtain a fast and accurate algorithm for the decomposition~\eqref{eqn:error-decomposition} for video surveillance data, we circumvent the problem using an orthogonalization trick similar to Gram Schimdt orthogonalization process~\cite{giraud2005loss}. In particular, between alternatively using Eq.~\eqref{eqn:estimating-eqn-1}-\eqref{eqn:estimating-eqn-3}, the estimates $\bb{a}$ and $\bb{b}$ are updated as 
\begin{equation}
    \bb{a} \leftarrow \bb{a} - \sum_{r = 1}^{k} {\bb{a}}\transpose \bb{a}_r, \ 
    \bb{b} \leftarrow \bb{b} - \sum_{r = 1}^{k} {\bb{b}}\transpose \bb{b}_r, \ 
    \label{eqn:GS-ortho}
\end{equation}
\noindent where the symbol $\leftarrow$ denotes the assignment operator, i.e., the value of left hand side of Eq.~\eqref{eqn:GS-ortho} is updated with the value of the right hand side. Note that, such an orthogonalization step need not be performed for the first singular value, but is performed only from the estimation of the subsequent singular values. Including the singular values $\lambda$ and restricting the vectors $\bb{a}$ and $\bb{b}$ to have unit $L_2$-norm, we ultimately consider the estimation of the parameter $\bb{\theta} = (\lambda, \bb{a}, \bb{b}, \sigma^2)$. The final algorithm has been outlined below in Figure~\ref{algo:rSVDdpd}.

\begin{figure}[!t]
    \begin{algorithmic}
        \REQUIRE Video frames $\bb{V}$ of dimension $h \times w \times p$, $0 < \alpha < 1$, background rank $r \geq 0$.
        \STATE Initialize matrix $\bb{X}_{hw \times p}$.
        \FOR{$i = 1$ to $p$}
            \STATE $\bb{X}[,i] \leftarrow \text{vec}(\bb{V}[,,i])$
        \ENDFOR
        \FOR{$k = 1$ to $r$}
            \STATE Let $\bb{a}_1^{\ast}, \bb{a}_2^{\ast}, \dots, \bb{a}_k^{\ast}$ and $\bb{b}_1^{\ast}, \bb{b}_2^{\ast}, \dots, \bb{b}_k^{\ast}$ are $k$ estimated singular vectors already obtained.
            \STATE Initialize $\lambda_{k+1}^{(0)}, \bb{a}_{k+1}^{(0)}, \bb{b}_{k+1}^{(0)}$ as first singular value and left and right singular vectors of $\bb{X}$.
            \STATE $t \leftarrow 0$.
            \REPEAT 
                \STATE $\bb{c} \leftarrow \bb{a}_{k+1}^{(t)}$
                \STATE $w_{ij} \leftarrow \exp\left(-\dfrac{(x_{ij} - c_ib_{(k+1)j}^{(t)})^2 }{2(\sigma^2)^{(t)}} \right)$
                \FOR{$i = 1, 2, \dots hw$}
                    \STATE $c_i \leftarrow \left(\dfrac{\sum_j w_{ij}^\alpha x_{ij} (b_{(k+1)j}^{(t)}) }{\sum_j w_{ij}^{\alpha} (b_{(k+1)j}^{(t)})^{2} } \right)$
                \ENDFOR
                \STATE $\bb{c} \leftarrow \bb{c} - \sum_{r=1}^{k} \bb{c}^{\top}\bb{a}_{r}$.
                \STATE $\lambda_{(k+1)}^{(t+1)} \leftarrow \Vert \bb{c} \Vert$.
                \STATE $\bb{a}_{k+1}^{(t+1)} \leftarrow \bb{c} / \lambda_{(k+1)}^{(t+1)}$
                \STATE $\bb{d} \leftarrow \bb{b}_{k+1}^{(t)}$
                \STATE $w_{ij} \leftarrow \exp\left(-\dfrac{(x_{ij} - a_{(k+1)i}^{(t+1)}d_j )^2 }{2(\sigma^2)^{(t)}} \right)$
                \FOR{$j = 1, 2, \dots p$}
                    \STATE $d_j \leftarrow \left(\dfrac{\sum_i w_{ij}^\alpha x_{ij} (a_{(k+1)i}^{(t+1)}) }{\sum_i w_{ij}^{\alpha} (a_{(k+1)i}^{(t+1)})^{2} } \right)$
                \ENDFOR
                \STATE $\bb{d} \leftarrow \bb{d} - \sum_{r=1}^{k} \bb{d}^{\top}\bb{b}_{r}$.
                \STATE $\lambda_{(k+1)}^{(t+1)} \leftarrow \Vert \bb{d} \Vert$.
                \STATE $\bb{b}_{k+1}^{(t+1)} \leftarrow \bb{d} / \lambda_{(k+1)}^{(t+1)}$.
                \STATE $e_{ij} \leftarrow x_{ij} - \lambda_{(k+1)}^{(t+1)} a_{(k+1)i}^{(t+1)}b_{(k+1)j}^{(t+1)}$
                \STATE $w_{ij} \leftarrow \exp(e_{ij})$
                \STATE $(\sigma^2)^{(t+1)} \leftarrow \dfrac{ \sum_i\sum_j w_{ij}^{\alpha} e_{ij}^2 }{\sum_i \sum_j w_{ij}^{\alpha} - (\alpha/(1 + \alpha)^{3/2})}$
                \STATE $t \leftarrow (t+1)$
            \UNTIL{convergence}
            \STATE $X \leftarrow X - \lambda_{(k+1)}^{(t+1)} \bb{a}_{(k+1)}^{(t+1)}(\bb{b}_{(k+1)}^{(t+1)})^{\top}$
        \ENDFOR
        \STATE \textbf{Output:} Background = $\sum_{k=1}^r \widehat{\lambda}_{k}^{(t+1)} \widehat{a}_k^{(t+1)} \left( \widehat{b}_k^{(t+1)} \right)^{\top}$ where each column is converted into an image of dimension $h \times w$.
    \end{algorithmic}
    \caption{rSVDdpd Background Modelling Algorithm}
    \label{algo:rSVDdpd}
\end{figure}

\subsection{Choice of hyperparameters}\label{sec:hyperparameters}

There are two hyperparameters to the rSVDdpd algorithm: the robustness parameter $\alpha \in [0,1]$ and the rank of the matrix $\bb{L}$ representing background content. The robustness parameter $\alpha$ in the objective function~\eqref{eqn:H-theta} provides a bridge between robustness and efficiency of estimation. Through extensive simulation studies, we have seen that higher values of the robustness parameter $\alpha$ tend to produce background estimates with smaller bias, under any kind of contamination. On the other side of the coin, the estimated pixel values of the background content exhibit higher variance with an increase in $\alpha$. To determine the optimal choice of $\alpha$, we consider a conditional MSE criterion as in~\cite{huang2009analysis}. Some elementary calculation yields that the optimal choice of the robustness parameter is the minimizer of the criterion 
\begin{multline*}
    (n + p) (\widehat{\sigma}^{(\alpha)})^2 \left( 1 + \dfrac{\alpha^2}{1 + 2\alpha} \right)^{3/2}\\
    + \dfrac{1}{r} \sum_{k=1}^r \Vert \widehat{\lambda}_k^{(\alpha)}\widehat{\boldsymbol{a}}_k^{(\alpha)} - \widehat{\lambda}_k^{(1)}\widehat{\boldsymbol{a}}_k^{(1)}\Vert_2^2 \\
    + \dfrac{1}{r}\sum_{k=1}^r \Vert \widehat{\lambda}_k^{(\alpha)}\widehat{\boldsymbol{b}}_k^{(\alpha)} - \widehat{\lambda}_k^{(1)}\widehat{\boldsymbol{b}}_k^{(1)}\Vert_2^2
\end{multline*}
\noindent where $\widehat{\lambda}_k^{(\alpha)}, \widehat{\boldsymbol{a}}_k^{(\alpha)}, \widehat{\bb{b}}_k^{(\alpha)}$ are the estimates of $k$-th singular value and vectors as obtained by the proposed rSVDdpd algorithm with robustness parameter $\alpha$.

The rank of $\bb{L}$, on the other hand, controls the amount of variation that will be allowed in the background content. For instance, a rank one approximation $\bb{L}$ will ensure that the columns of $\bb{L}$ are scalar multiples of each other, i.e., the pixels values of the background from one frame to other change multiplicatively. This is true if there are only illumination changes in the background, from day to night, from clear sky to overcast or fog. However, in presence of slight movements of the trees and shadows, the assumption of the rank one approximation does not hold true. In such cases, the background content forms a matrix $\bb{L}$ with rank two or more. Thus, to determine the rank of the matrix $\bb{L}$, we start with the usual singular value decomposition of $\bb{L}$ and choose the rank $r$ such that the first $r$ singular values and vectors together explain $(1-\epsilon)$ proportion of the variation, with common choices of $\epsilon$ being $0.1$ or $0.25$. Therefore, we extract the first $r^\ast$ singular values and vectors of $\bb{X}$ using rSVDdpd algorithm where
\begin{equation*}
    r^\ast = \min\left\{ 1 \leq r \leq \min(n,p): \dfrac{\sum_{k=1}^{r} (\widehat{\lambda}^{(0)}_k)^2}{\sum_{k=1}^{\min(n,p)} (\widehat{\lambda}^{(0)}_k)^2  } > (1-\epsilon) \right\}
\end{equation*}
\noindent where $\widehat{\lambda}^{(0)}_k$ is the $k$-th singular value of $\bb{X}$ as estimated by usual SVD method. The usage of the usual nonrobust SVD to determine this rank is purely from a computational point of view, which has been popularly used by several authors~\cite{he2012incremental,xu2012robust}. One may also consider bi-cross validation approach~\cite{owen2009bi} for estimating the rank $r^\ast$ subject to the available computational constraints.

\subsection{Convergence and Mathematical Properties}\label{sec:math-property}

The convergence of the rSVDdpd algorithm based on the alternating iterative equations~\eqref{eqn:algo-eqn} can be guaranteed under fairly minimal assumptions about the entries of the data matrix $\bb{X}$. First, rewrite the objective function $H_{n,p}(\bb{a}, \bb{b}, \sigma^2)$ for estimating the rank one decomposition as 
\begin{multline}
    \Hcal_{n,p}(\lambda, \bb{u}, \bb{v}, \sigma^2)
    = (2\pi)^{-\alpha/2}\sigma^{-\alpha} \left[ (1+\alpha)^{-1/2} \right. \\
    \left. \dfrac{(1+\alpha)}{\alpha} \dfrac{1}{np}\sum_{i=1}^n \sum_{j=1}^p \exp\left\{ - \alpha\dfrac{(x_{ij} - \lambda a_i b_j)^2}{2\sigma^2} \right\} \right]
    \label{eqn:Hcal-function}
\end{multline}
\noindent where $\lambda = \Vert \bb{a}\Vert \Vert \bb{b}\Vert$ is the first singular value and $\bb{u}$ and $\bb{v}$ are the normalized unit vectors obtained from $\bb{a}$ and $\bb{b}$ respectively. For fixed dimensions $n$ and $p$, the function $\Hcal$ is continuous in all of its arguments. The unit vectors $\bb{u}$ and $\bb{v}$ both lie on the compact set, namely on $(n-1)$ and $(p-1)$-dimensional unit hyperspheres respectively, therefore, for fixed $\lambda$ and $\sigma^2$, the objective function $\Hcal$ is bounded below. Now, as $\lambda$ tends to $0$ or to $\infty$, in both cases $\Hcal$ remains bounded. And, $\Hcal$ also remains bounded below as the scale parameter $\sigma^2$ decreases to $0$ or increases to $\infty$. Therefore, combining all of these pieces of information, one can conclude that $\Hcal$ must necessarily be bounded below, hence must attain a minimum. Then the convergence of the iteration rules~\eqref{eqn:algo-eqn} easily follows from observing that each such iteration must necessarily decrease the value of the objective function~\eqref{eqn:Hcal-function} and an application of Bolzano-Weierstrass theorem. In particular, since the values of $x_{ij}$ will be bounded for the video surveillance application, the assurance for convergence can be made uniform over the choice of $n$ and $p$; see Roy et al.~\cite{roy2021new} for the technical details.

The converged rSVDdpd estimator has also various desirable theoretical properties as discussed in detail by Roy et al.~\cite{roy2021new}. Let us denote the theoretical minimizer of Eq.~\eqref{eqn:H-theta} as $\boldsymbol{\theta}^g = (\lambda^g, \{ a_i^g \}_{i = 1}^{n}, \{ b_j^g \}_{j=1}^{p}, (\sigma^g)^2 )$ satisfying
\begin{equation}
    \begin{split}
        \sum_{i=1}^{n} (a_i^g)^2 & = \sum_{j=1}^{p} (b_j^g)^2 = 1, \\
        \lambda^g a_i^g & = \argmin_{a} \int V(x; a, b_j^g, (\sigma^g)^2 ) g_{ij}(x) dx,\\
        \lambda^g b_j^g & = \argmin_{b} \int V(x; a_i^g, b, (\sigma^g)^2 ) g_{ij}(x)dx, \\
        (\sigma^g)^2 & = \argmin_{\sigma^2}  \int V(x; \lambda^g a_i^g, b_j^g, \sigma^2) g_{ij}(x)dx.
    \end{split}
\end{equation}
\noindent where $g_{ij}$ is the true density function of $x_{ij}$ representing the statistical model for the background. Then, the following properties are proved for the converged rSVDdpd estimators; see Roy et al.~\cite{roy2021new} for their proofs and other technical details.
\begin{enumerate}
    \item The converged rSVDdpd estimator is equivariant under scale and permutation transformations like the usual singular values and vectors.
    \item The converged rSVDdpd estimator is consistent for $\boldsymbol{\theta}^g$ as $n$ and $p$ increases to infinity.
\end{enumerate}

\begin{figure}[!t]
    \centering
    \includegraphics[width =3in]{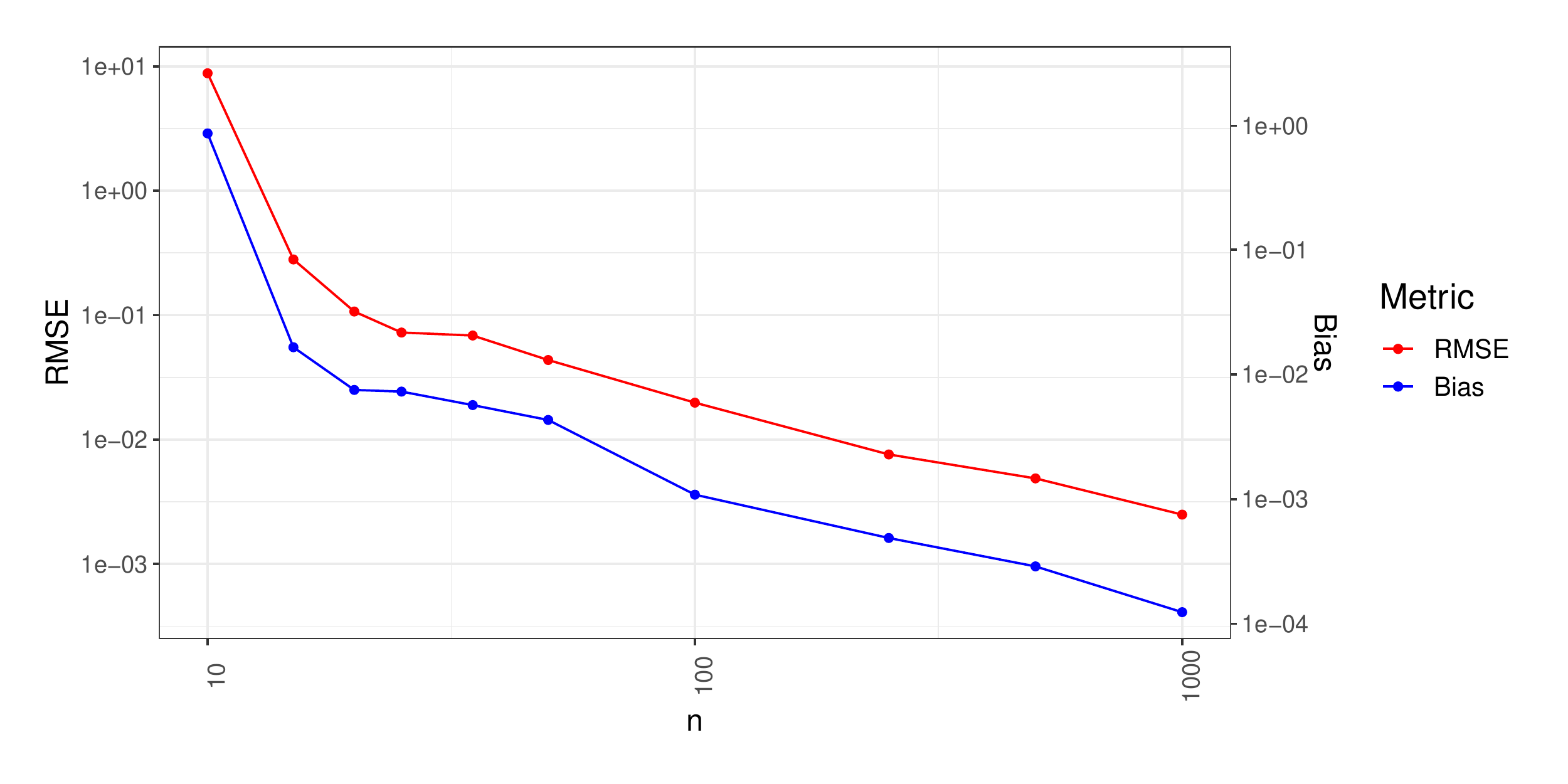}
    \caption{Empirical Bias and RMSE of the first singular value of an $n \times n$ matrix comprising a rank three matrix and an error matrix (Both the horizontal and vertical axes are in logarithmic scale)}
    \label{fig:consistency}
\end{figure}

Figure~\ref{fig:consistency} depicts the empirical bias and the root-mean-squared error (RMSE) of the first singular value of an $n \times n$ matrix $\bb{X} = \bb{L} + \bb{E}$, where $\bb{L}$ is a matrix with rank three and $\bb{E}$ is a matrix with i.i.d. entries generated from a Gaussian distribution with mean $0$ and standard deviation $1/n$. This also demonstrates the consistency of the rSVDdpd estimator by empirically verifying that both the bias and the RMSE of the first singular value tend to $0$ as the dimension $n$ tends to infinity.

\section{Experimental Results}\label{sec:experiments}

\subsection{Experimental Setup}

As explained in Section~\ref{sec:introduction}, we focus on comparing the proposed rSVDdpd method primarily with the existing robust PCA algorithms for background modelling. As most of the recent advancements in the robust PCA based approaches concern themselves with reducing the computational complexity of the optimization algorithm rather than significantly improving the performance of the background modelling estimates, we restrict our attention to the major robust PCA approaches and the corresponding seminal algorithms. These include the exact Principal Component Pursuit (PCP) via Augmented Lagrangian Method (ALM)~\cite{candes2011}, inexact ALM via alternating direction~\cite{shen2014augmented}, Outlier Pursuit (OP)~\cite{xu2012robust}, Sparse Regularized PCP (SRPCP)~\cite{liu2017sparsity} and Variational Bayesian method (VB)~\cite{babacan2012sparse}. In addition, we also consider one statistical algorithm Go Decomposition (GoDec)~\cite{zhou2011godec} and an incremental algorithm, namely the Grassmannian Robust Adaptive Subspace Tracking Algorithm (GRASTA)~\cite{he2011online,he2012incremental}. While the first set of algorithms work with a batch of multiple frames at a time to extract the foreground content, the last algorithm GRASTA is online in nature, it processes one frame of the video at a time and uses the information from the previous frames to extract foreground of the current video frame. R implementation of rSVDdpd and exact PCP method for robust PCA are available in the rSVDdpd~\cite{rSVDdpdR} and the RPCA~\cite{rpcaR} packages, respectively. The MATLAB implementations of the rest of the algorithms are available in Background Subtraction Website\footnote{\url{https://sites.google.com/site/backgroundsubtraction/Home}} which have been converted to R implementation to compare these methods on a level ground. All the experiments have been done in an Azure virtual machine with 4 vCPUs and 32 GB of RAM.

To compare these methods and measure performances, we consider the Background Models Challenge benchmark dataset~\cite{vacavant2012benchmark}. This benchmark dataset has also been used in earlier studies to provide a detailed comparison of several background modelling algorithms~\cite{bouwmans2014robust}. The dataset contains $20$ synthetic videos with ground truth foreground masks. The synthetic videos are generated by SIVIC software and closely resemble different kinds of natural tampering present in real-life video surveillance data such as noises, artifacts, shadows, change of illumination, presence of fog and rain, etc. The whole set of videos comprises to two kinds of backgrounds, one of a straight street highway (Street), while the other contains the simulated video footage of a traffic junction (Rotary). Among these $10$ pairs of videos, we use the first of each pair to train the parameters of these algorithms and use the second of each pair to test the performance of the respective trained models. Due to the computational limitations of the existing algorithms, we train each of the algorithms using batches of $120$ frames from the videos at a time and find the optimal parameter with the best $F1$-measure. The $F1$-measure is the harmonic mean of the precision and recall, which is generally accepted as a robust measure of the performance of any classification algorithm. Using the ground truth mask provided in the dataset, we calculate the precision as the ratio of the correctly predicted foreground pixels and the predicted foreground pixels, while the recall is calculated as the ratio of the correctly predicted foreground pixels and the number of pixels assigned to the foreground content in the ground truth video.

\subsection{Discussions of Results for the Benchmark Dataset}

\begin{table}[!t]    
    \caption{Time taken (in seconds) for processing a single frame for different background modelling algorithms\\ (Algorithms are in descending order of speed)}
    \label{tab:time-comparison}
    \begin{tabular}{lll}
    \toprule
    Method & Average Time  & Std. dev. \\ 
    & (in seconds/frame) & (in seconds/frame)\\
    \midrule
     GRASTA~\cite{he2012incremental} & 0.37 & 0.01 \\ 
     rSVDdpd (ours) & 2.86 & 0.84 \\ 
     VB~\cite{babacan2012sparse} & 7.06 & 1.39 \\ 
     GoDec~\cite{zhou2011godec} & 12.06 & 0.29 \\ 
     OP~\cite{xu2012robust} & 23.42 & 0.22 \\ 
     ALM~\cite{shen2014augmented} & 34.85 & 7.40 \\ 
    SRPCP~\cite{liu2017sparsity} & 54.25 & 9.12 \\  
    RPCA~\cite{candes2011} & 136.41 & 27.26 \\      
    \bottomrule
\end{tabular}
\end{table}

\begin{table*}[!t]
    \caption{Benchmark Results for different variants of the Street Video (Video number Vi1j denotes $i$-th variant of the Street Video and $j = 1$ or $2$ denote the training and the testing video respectively); In each column, bold symbols indicate the best performing algorithm with respect to the $F_1$ measure.}
    \label{tab:benchmark-result-street}
    \centering
    \begin{tabular}{@{}llllllllllll@{}}
        \toprule
        Method & Measures  & V111 & V112 & V211 & V212 & V311 & V312 & V411 & V412 & V511 & V512 \\ \midrule
        \multirow{3}{*}{RPCA~\cite{candes2011}}
        & Precision & 0.766 & 0.741 & 0.788 & 0.764 & 0.751 & 0.761 & 0.983 & 0.784 & 0.617 & 0.487\\
        & Recall & 0.768 & 0.767 & 0.747 & 0.744 & 0.7 & 0.698 & 0.655 & 0.675 & 0.568 & 0.49\\
        & F1 & 0.767 & 0.754 & 0.767 & 0.754 & 0.725 & 0.728 & 0.787 & 0.726 & 0.592 & 0.489\\
        \midrule
        \multirow{3}{*}{ALM~\cite{shen2014augmented}}
        & Precision & 0.766 & 0.741 & 0.811 & 0.765 & 0.752 & 0.762 & 0.984 & 0.785 & 0.626 & 0.486\\
        & Recall & 0.768 & 0.767 & 0.729 & 0.745 & 0.7 & 0.698 & 0.655 & 0.675 & 0.568 & 0.49\\
        & F1 & 0.767 & 0.754 & 0.768 & 0.755 & 0.725 & 0.729 & 0.787 & 0.726 & 0.595 & 0.488\\
        \midrule
        \multirow{3}{*}{SRPCP~\cite{liu2017sparsity}}
        & Precision & 0.766 & 0.741 & 0.811 & 0.765 & 0.752 & 0.762 & 0.984 & 0.785 & 0.626 & 0.486\\
        & Recall & 0.768 & 0.767 & 0.729 & 0.745 & 0.7 & 0.698 & 0.657 & 0.675 & 0.568 & 0.49\\
        & F1 & 0.767 & 0.754 & 0.768 & 0.755 & 0.725 & 0.729 & \textbf{0.788} & 0.726 & 0.595 & 0.488\\
        \midrule
        \multirow{3}{*}{VB~\cite{babacan2012sparse}}
        & Precision & 0.774 & 0.749 & 0.809 & 0.764 & 0.75 & 0.761 & 0.985 & 0.781 & 0.323 & 0.38\\
        & Recall & 0.767 & 0.766 & 0.729 & 0.745 & 0.701 & 0.698 & 0.654 & 0.649 & 0.48 & 0.435\\
        & F1 & 0.77 & 0.757 & 0.767 & 0.754 & 0.725 & 0.728 & 0.786 & 0.709 & 0.386 & 0.405\\
        \midrule
        \multirow{3}{*}{OP~\cite{xu2012robust}}
        & Precision & 0.783 & 0.776 & 0.825 & 0.794 & 0.757 & 0.837 & 0.985 & 0.797 & 0.686 & 0.619\\
        & Recall & 0.769 & 0.748 & 0.727 & 0.723 & 0.699 & 0.532 & 0.653 & 0.643 & 0.601 & 0.5\\
        & F1 & 0.776 & 0.762 & 0.773 & \textbf{0.757} & \textbf{0.727} & 0.65 & 0.785 & 0.712 & 0.641 & \textbf{0.553}\\
        \midrule
        \multirow{3}{*}{GoDec~\cite{zhou2011godec}}
        & Precision & 0.844 & 0.8 & 0.786 & 0.762 & 0.749 & 0.771 & 0.981 & 0.51 & 0.205 & 0.414\\
        & Recall & 0.686 & 0.676 & 0.748 & 0.745 & 0.701 & 0.595 & 0.654 & 0.443 & 0.364 & 0.214\\
        & F1 & 0.757 & 0.733 & 0.767 & 0.754 & 0.724 & 0.672 & 0.785 & 0.474 & 0.263 & 0.282\\
        \midrule
        \multirow{3}{*}{GRASTA~\cite{he2012incremental}}
        & Precision & 0.646 & 0.766 & 0.613 & 0.758 & 0.764 & 0.89 & 0.987 & 0.165 & 0.338 & 0.374\\
        & Recall & 0.764 & 0.644 & 0.727 & 0.712 & 0.613 & 0.282 & 0.652 & 0.189 & 0.395 & 0.367\\
        & F1 & 0.700 & 0.699 & 0.665 & 0.734 & 0.68 & 0.428 & 0.785 & 0.176 & 0.364 & 0.371\\        
        \midrule
        \multirow{3}{*}{rSVDdpd ((ours)}
        & Precision & 0.783 & 0.776 & 0.78 & 0.769 & 0.753 & 0.842 & 0.987 & 0.8 & 0.785 & 0.626\\
        & Recall & 0.772 & 0.768 & 0.769 & 0.746 & 0.699 & 0.683 & 0.653 & 0.735 & 0.601 & 0.497\\
        & F1 & \textbf{0.777} & \textbf{0.772} & \textbf{0.774} & \textbf{0.757} & 0.725 & \textbf{0.754} & 0.787 & \textbf{0.766} & \textbf{0.681} & \textbf{0.553}\\
        \bottomrule
    \end{tabular}
\end{table*}

\begin{table*}[!t]
\caption{Benchmark Results for different variants of the Rotary Video (Video number Vi2j denotes $i$-th variant of the Rotary Video and $j = 1$ or $2$ denote the training and the testing video respectively); In each column, bold symbols indicate the best performing algorithm with respect to the $F_1$ measure.}
\label{tab:benchmark-result-rotary}
\centering
\begin{tabular}{@{}llllllllllll@{}}
    \toprule
    Method & Measures  & V121 & V122 & V221 & V222 & V321 & V322 & V421 & V422 & V521 & V522 \\ 
    \midrule
    \multirow{3}{*}{RPCA~\cite{candes2011}}
    & Precision & 0.771 & 0.771 & 0.792 & 0.795 & 0.781 & 0.781 & 0.893 & 0.698 & 0.425 & 0.496\\
    & Recall & 0.771 & 0.769 & 0.733 & 0.738 & 0.656 & 0.659 & 0.612 & 0.547 & 0.592 & 0.599\\
    & F1 & 0.771 & 0.77 & 0.761 & 0.766 & 0.713 & 0.714 & 0.726 & 0.614 & 0.495 & 0.542\\
    \midrule
    \multirow{3}{*}{ALM~\cite{shen2014augmented}}
    & Precision & 0.772 & 0.771 & 0.792 & 0.796 & 0.781 & 0.781 & 0.898 & 0.698 & 0.422 & 0.491\\
    & Recall & 0.771 & 0.769 & 0.733 & 0.738 & 0.656 & 0.659 & 0.611 & 0.546 & 0.592 & 0.599\\
    & F1 & 0.772 & 0.77 & 0.761 & 0.766 & 0.713 & 0.714 & 0.727 & 0.613 & 0.493 & 0.54\\
    \midrule
    \multirow{3}{*}{SRPCP~\cite{liu2017sparsity}}
    & Precision & 0.772 & 0.771 & 0.792 & 0.796 & 0.781 & 0.781 & 0.898 & 0.698 & 0.422 & 0.491\\
    & Recall & 0.771 & 0.769 & 0.733 & 0.738 & 0.656 & 0.659 & 0.611 & 0.546 & 0.592 & 0.599\\
    & F1 & 0.772 & 0.77 & 0.761 & 0.766 & 0.713 & \textbf{0.715} & 0.727 & 0.613 & 0.493 & 0.54\\
    \midrule
    \multirow{3}{*}{VB~\cite{babacan2012sparse}}
    & Precision & 0.772 & 0.772 & 0.793 & 0.797 & 0.776 & 0.773 & 0.904 & 0.683 & 0.447 & 0.483\\
    & Recall & 0.771 & 0.769 & 0.733 & 0.738 & 0.658 & 0.661 & 0.611 & 0.518 & 0.526 & 0.527\\
    & F1 & 0.772 & 0.77 & \textbf{0.762} & 0.766 & \textbf{0.712} & 0.713 & 0.729 & 0.589 & 0.483 & 0.504\\
    \midrule
    \multirow{3}{*}{OP~\cite{xu2012robust}}
    & Precision & 0.806 & 0.79 & 0.792 & 0.802 & 0.751 & 0.797 & 0.883 & 0.655 & 0.551 & 0.544\\
    & Recall & 0.728 & 0.733 & 0.713 & 0.7 & 0.657 & 0.468 & 0.611 & 0.517 & 0.467 & 0.488\\
    & F1 & 0.765 & 0.76 & 0.75 & 0.749 & 0.701 & 0.59 & 0.722 & 0.577 & 0.506 & 0.514\\
    \midrule
    \multirow{3}{*}{GoDec~\cite{zhou2011godec}}
    & Precision & 0.815 & 0.795 & 0.758 & 0.762 & 0.737 & 0.796 & 0.987 & 0.57 & 0.502 & 0.512\\
    & Recall & 0.624 & 0.688 & 0.734 & 0.739 & 0.661 & 0.539 & 0.57 & 0.587 & 0.476 & 0.5\\
    & F1 & 0.707 & 0.738 & 0.746 & 0.75 & 0.697 & 0.642 & 0.723 & 0.578 & 0.488 & 0.506\\
    \midrule
    \multirow{3}{*}{GRASTA~\cite{he2012incremental}}
    & Precision & 0.82 & 0.56 & 0.604 & 0.797 & 0.758 & 0.762 & 0.892 & 0.45 & 0.334 & 0.346\\
    & Recall & 0.684 & 0.692 & 0.69 & 0.7 & 0.628 & 0.301 & 0.61 & 0.283 & 0.603 & 0.611\\
    & F1 & 0.746 & 0.619 & 0.644 & 0.748 & 0.687 & 0.432 & 0.724 & 0.347 & 0.43 & 0.442\\
    \midrule
    \multirow{3}{*}{rSVDdpd (ours)}
    & Precision & 0.772 & 0.784 & 0.798 & 0.805 & 0.735 & 0.685 & 0.901 & 0.697 & 0.742 & 0.634\\
    & Recall & 0.792 & 0.769 & 0.729 & 0.733 & 0.686 & 0.718 & 0.65 & 0.58 & 0.56 & 0.586\\
    & F1 & \textbf{0.781} & \textbf{0.776} & \textbf{0.762} & \textbf{0.767} & 0.710 & 0.702 & \textbf{0.755} & \textbf{0.633} & \textbf{0.638} & \textbf{0.609}\\
    \bottomrule
\end{tabular}
\end{table*}

\begin{figure*}[!t]
    \centering
    \subfloat[]{\includegraphics[width=0.18\linewidth]{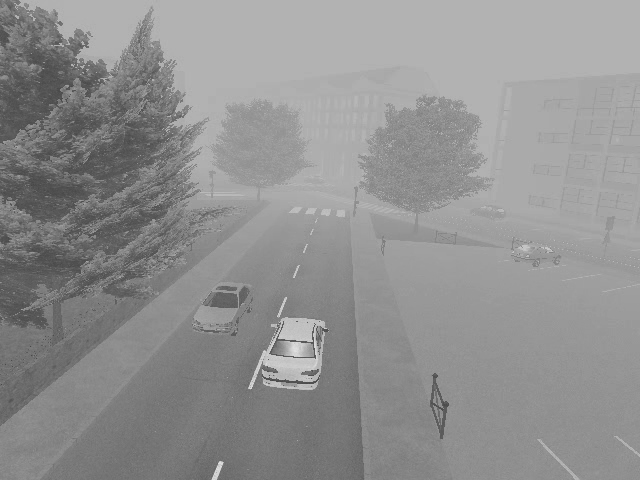}}
    \hfill
    \subfloat[]{\includegraphics[width=0.18\linewidth]{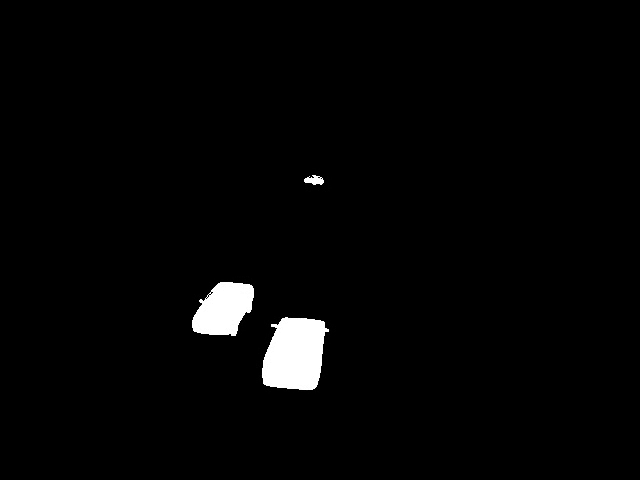}}
    \hfill
    \subfloat[]{\includegraphics[width=0.18\linewidth]{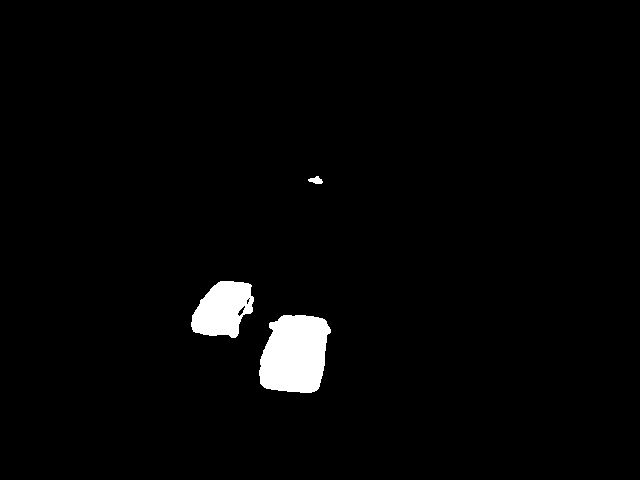}}
    \hfill
    \subfloat[]{\includegraphics[width=0.18\linewidth]{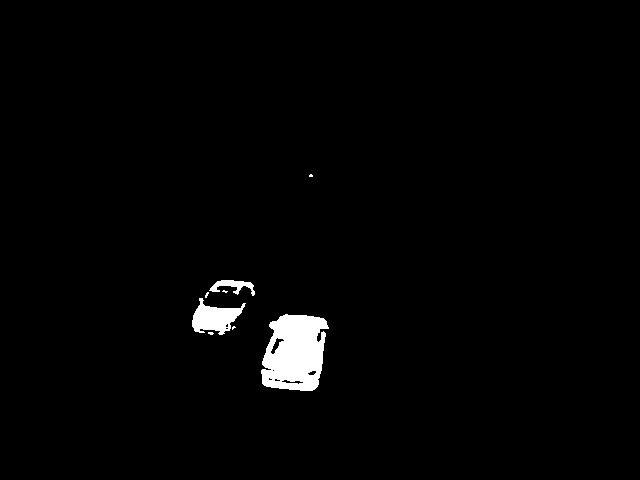}}
    \hfill
    \subfloat[]{\includegraphics[width=0.18\linewidth]{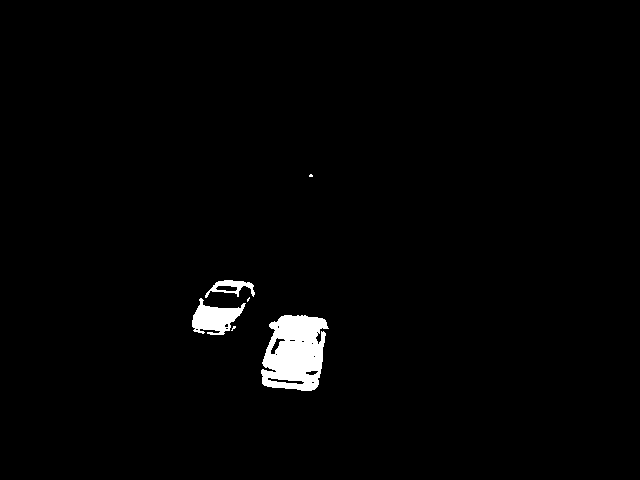}}
    \hfill
    \subfloat[]{\includegraphics[width=0.18\linewidth]{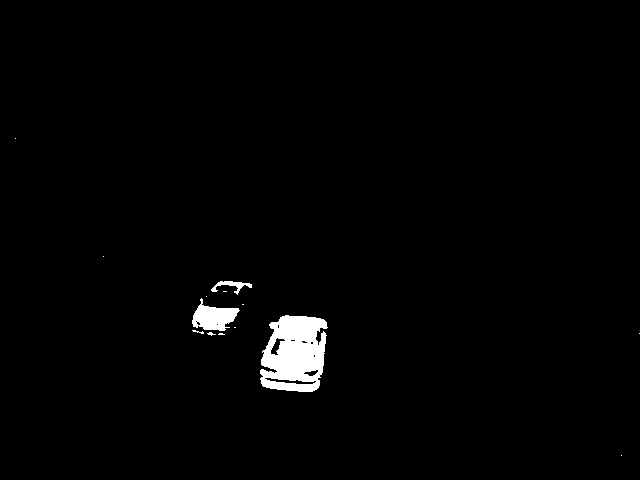}}
    \hfill
    \subfloat[]{\includegraphics[width=0.18\linewidth]{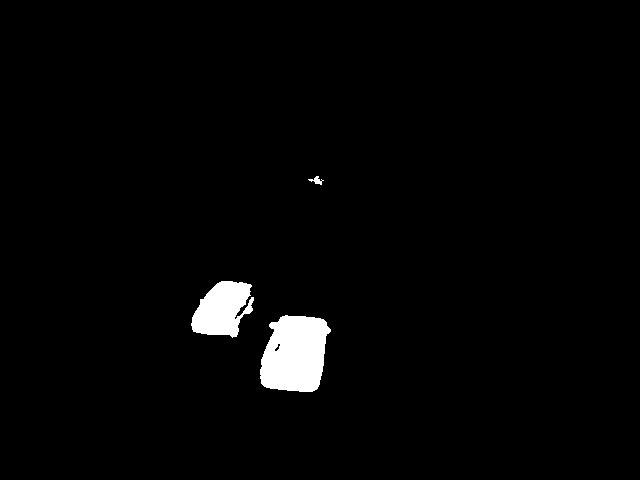}}
    \hfill
    \subfloat[]{\includegraphics[width=0.18\linewidth]{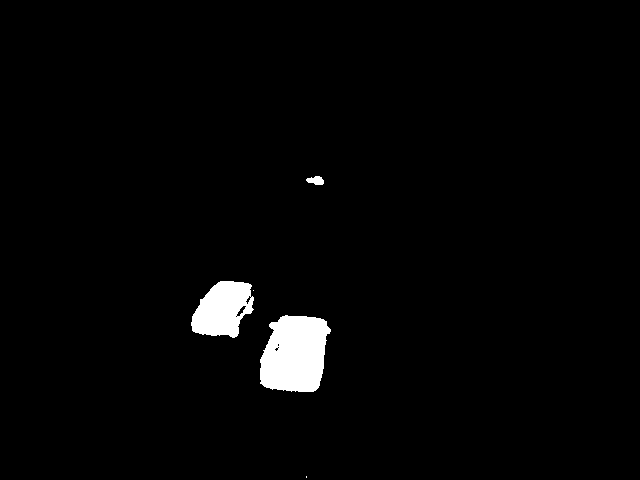}}
    \hfill
    \subfloat[]{\includegraphics[width=0.18\linewidth]{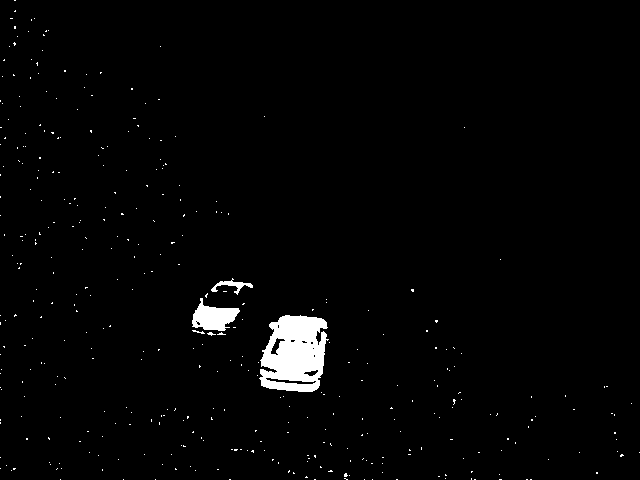}}
    \hfill
    \subfloat[]{\includegraphics[width=0.18\linewidth]{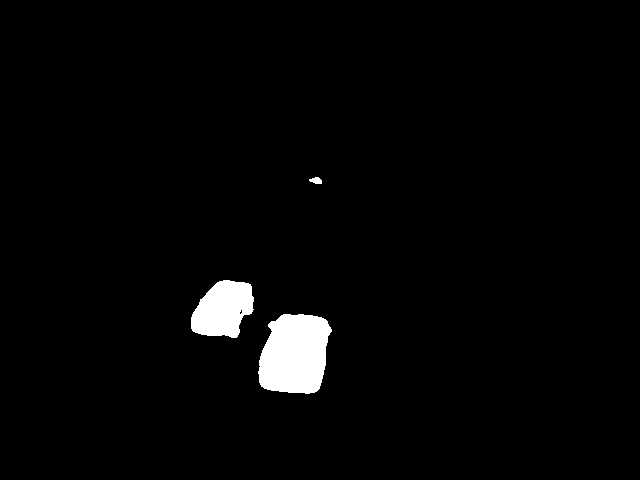}}
    \caption{One frame of the street video (a), ground truth (b) and estimated foreground mask from exact RPCA (c), inexact ALM (d), SRPCP (e), Variational Bayes (f), Outlier Pursuit (g), GoDec (h), GRASTA (i) and rSVDdpd (j) algorithms. Enlarged better quality images are available online at https://subroy13.github.io/rsvddpd-home/.}
    \label{fig:video-street}
\end{figure*}

\begin{figure*}[!t]
    \centering
    \subfloat[]{\includegraphics[width=0.18\linewidth]{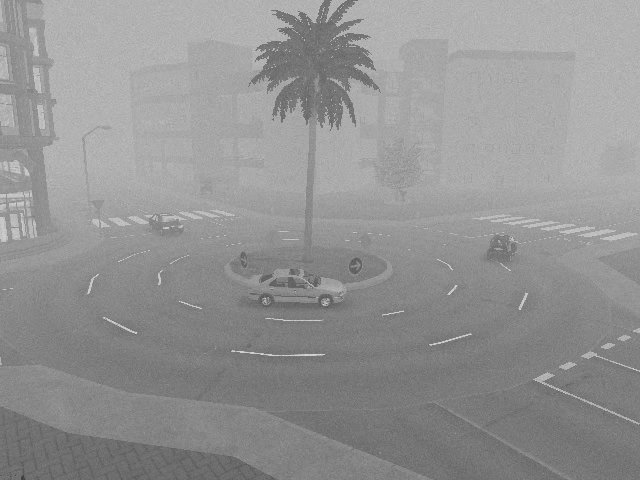}}
    \hfill
    \subfloat[]{\includegraphics[width=0.18\linewidth]{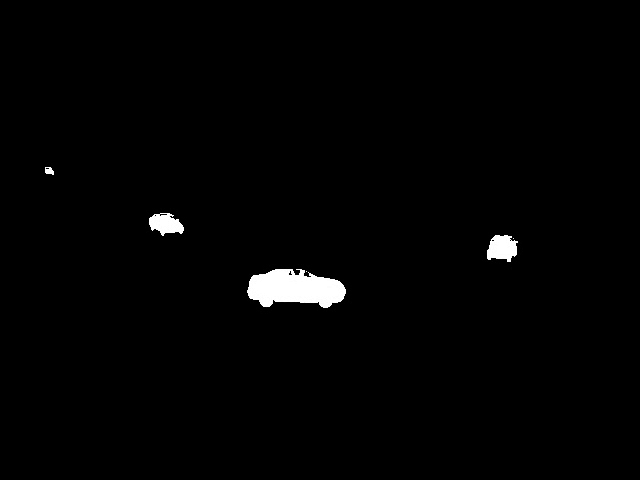}}
    \hfill
    \subfloat[]{\includegraphics[width=0.18\linewidth]{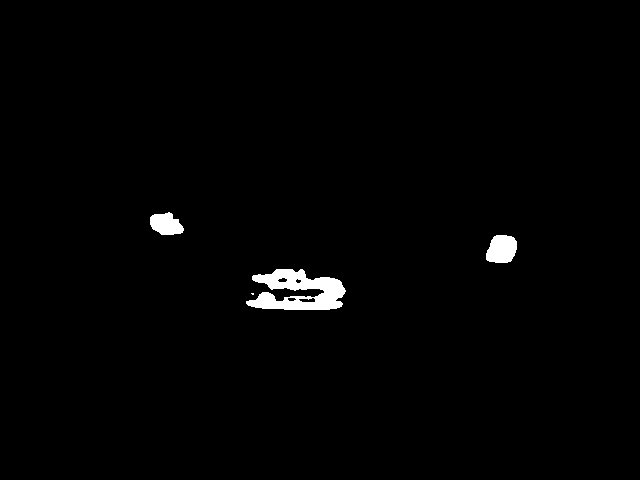}}
    \hfill
    \subfloat[]{\includegraphics[width=0.18\linewidth]{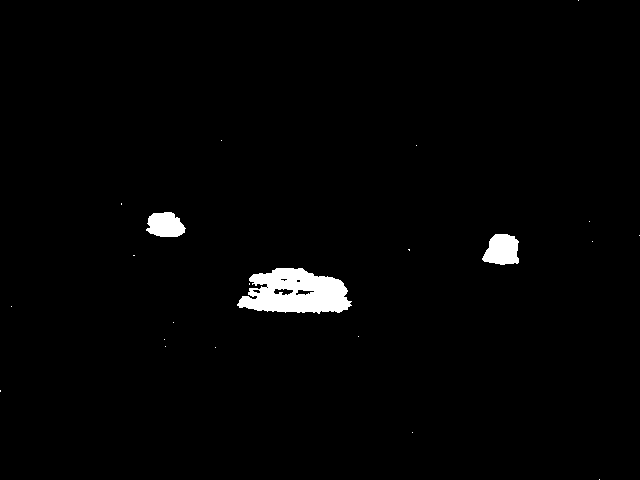}}
    \hfill
    \subfloat[]{\includegraphics[width=0.18\linewidth]{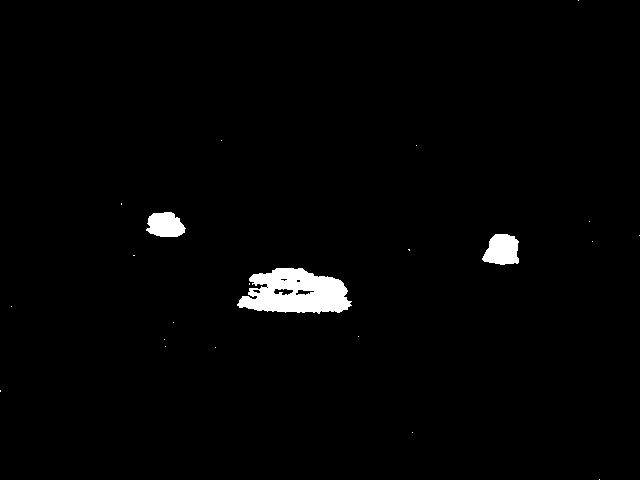}}
    \hfill
    \subfloat[]{\includegraphics[width=0.18\linewidth]{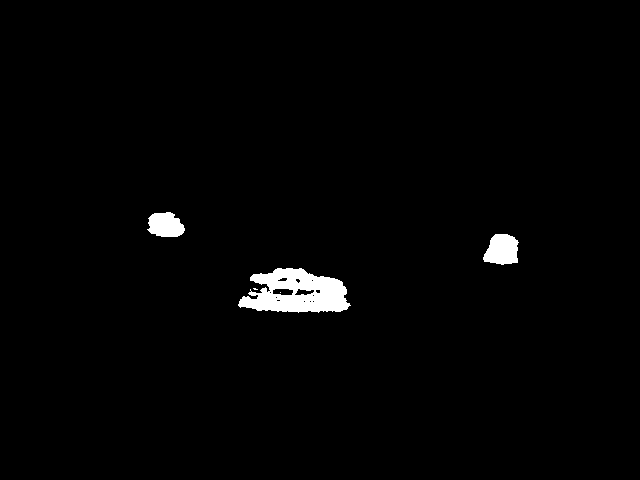}}
    \hfill
    \subfloat[]{\includegraphics[width=0.18\linewidth]{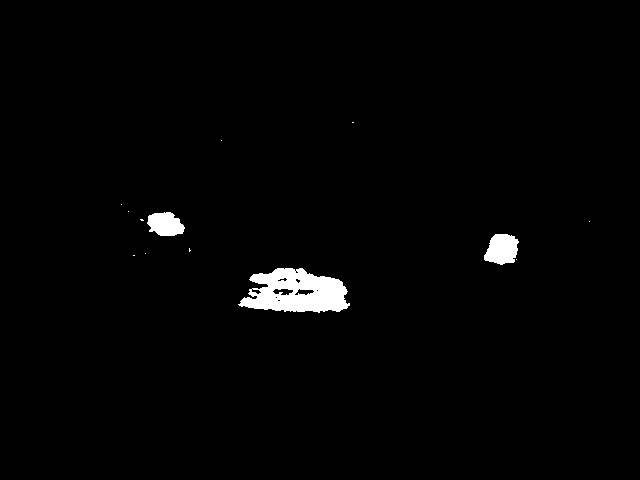}}
    \hfill
    \subfloat[]{\includegraphics[width=0.18\linewidth]{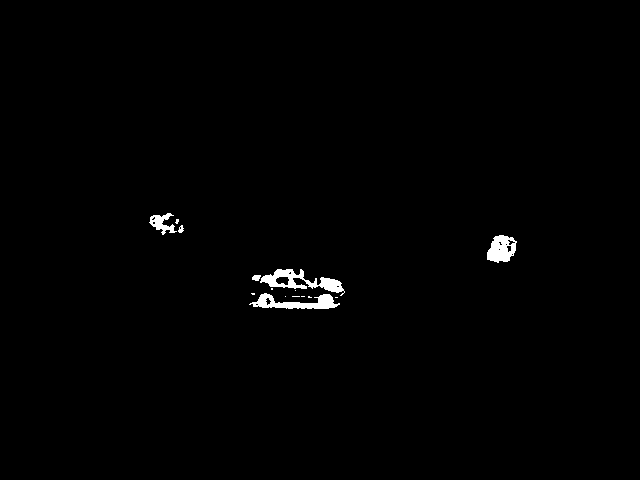}}
    \hfill
    \subfloat[]{\includegraphics[width=0.18\linewidth]{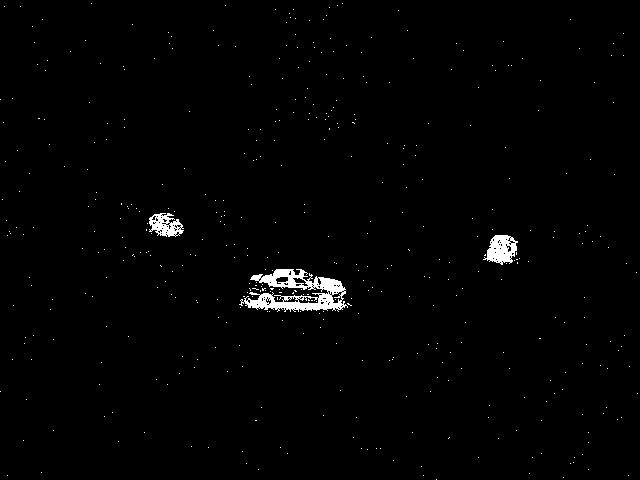}}
    \hfill
    \subfloat[]{\includegraphics[width=0.18\linewidth]{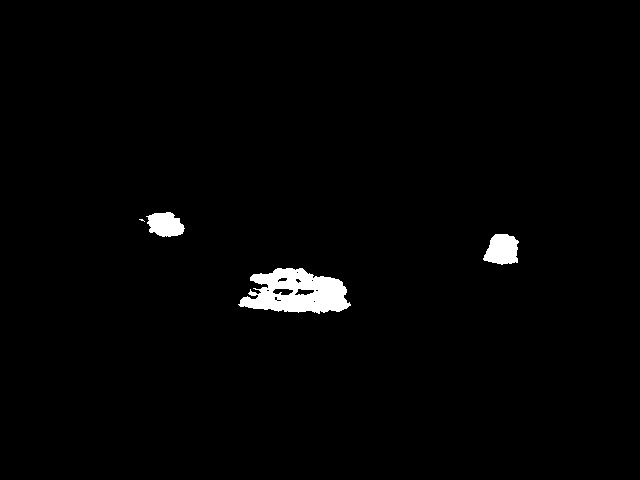}}
    \caption{One frame of the rotary video (a), ground truth (b) and estimated foreground mask from exact RPCA (c), inexact ALM (d), SRPCP (e), Variational Bayes (f), Outlier Pursuit (g), GoDec (h), GRASTA (i) and rSVDdpd (j) algorithms. Enlarged better quality images are available online at https://subroy13.github.io/rsvddpd-home/.}
    \label{fig:video-rotary}
\end{figure*}

In Table~\ref{tab:time-comparison}, we summarize the average and the standard deviation of the time taken per frame (in seconds) by each of the algorithms to converge and separate the background and foreground content. Only GRASTA and rSVDdpd algorithms complete the foreground extraction task in a reasonable amount of time to be useful in real-time video surveillance and object tracking. In contrast, the exact robust PCA method takes an enormous amount of time to converge, more than $2$ minutes per frame, thus destroying its practical utility. The variants of robust PCA, i.e., inexact ALM and SRPCP methods reduce the computational time significantly, but they are still not on par with the speed of rSVDdpd or GRASTA algorithms. This computational performance gain of rSVDdpd can be further boosted through the highly parallelizable structure of the iteration rules~\eqref{eqn:algo-eqn}, where each component $a_i$ (or $b_j$) of the singular vectors can be updated simultaneously. This empirically demonstrates how rSVDdpd solves the problem of scalability of the video surveillance background modelling methods and achieves high parallelizability.

In Table~\ref{tab:benchmark-result-street} and Table~\ref{tab:benchmark-result-rotary}, we summarize the performances of the different algorithms for foreground detection on the benchmark datasets. Clearly, the proposed rSVDdpd algorithm outperforms all the existing algorithms under consideration for almost all the benchmark videos. For some of the videos where rSVDdpd does not achieve the best $F1$-measure, the difference in $F1$-measure compared to the best output is too small to visually distinguish the extracted foregrounds. It turns out that the inexact ALM method and the SRPCP method have similar performances as the exact robust PCA algorithm, while the OP and VB methods perform better than these only in some specific scenarios, in particular when the background shows some continuous natural movement of the trees. Both GoDec and GRASTA algorithms do not work well under these situations. In comparison, the rSVDdpd algorithm is able to tackle all these scenarios by allowing the background matrix $\bb{L}$ to be of rank more than one. 

As seen from the above discussions, the rSVDdpd algorithm achieves superior or near-optimal compared to the existing robust PCA and background modelling methods by spending only a fraction of computational time. GRASTA, the only algorithm that is faster than the proposed rSVDdpd (indeed, faster than all other algorithms compared), achieves this speed at the cost of a very poor performance under different kinds of noises present in the real-life video data. A fast algorithm with little performance guarantee can only be of limited use. For instance, GRASTA performs worse than all the algorithms compared for video 412 and 422 of Table~\ref{tab:benchmark-result-street} and~\ref{tab:benchmark-result-rotary} respectively. These two videos have small amount of noises added throughout the video and fog and haze appear in between for some consecutive frames as the outlying observations. While the improvement (rSVDdpd  over GRASTA) is not necessarily absolute in all the other cases, many other videos (such as 312, 511, 512 in Table 2 and 322, 521, 522 in Table 3) show that the improvement is substantial in most cases. And in each of these cases the rSVDdpd is either the top performer, or is in the neighborhood of the best. The degree may vary by small amounts in the other videos, but the general pattern holds true. The rSVDdpd holds its own even when practically all the other algorithms show a largely degraded performance, e.g., as in the case of video 512. Also, rSVDdpd is the second best in our examined set in terms of speed (see Table~\ref{tab:time-comparison}), clearly beats the GRASTA algorithm in terms of performance accuracy, is competitive or better than all the other algorithms in terms of performance accuracy, and is 3 to 50 times faster than all the other algorithms compared except GRASTA. Thus, the proposed rSVDdpd keeps an optimal balance between speed and accuracy without losing much in both of these aspects.

Due to space constraint, we demonstrate the estimated foreground masks as obtained from only two frames of the two videos, using different existing algorithms and the proposed rSVDdpd algorithm in Figures~\ref{fig:video-street} and~\ref{fig:video-rotary}. As shown in these figures, both the exact robust PCA method and the rSVDdpd method lead to visually indistinguishable results; however rSVDdpd achieves the same at the cost of a significantly reduced computational time compared to the exact robust PCA algorithm. The inexact ALM, SRPCP and VB methods, which converge faster than the exact robust PCA method, sometimes find it difficult to extract the foreground object in the presence of clouds and fog. The proposed rSVDdpd algorithm can efficiently work in all these situations by considering a rank-one approximation of the background matrix $\bb{L}$. Due to the use of the robust MDPDE, any moving foreground objects are also captured through the rSVDdpd algorithm even when the cloud or the fog blends the foreground's grayscale intensity with the background's grayscale intensity. The other alternative competing methods for video surveillance, namely GRASTA and GoDec estimate a noisy background in general. Although these noises remain imperceptible at first, they result in a more noisy foreground, and such noisy pixels sometimes become visible in the thresholded foreground mask. As demonstrated earlier in Figure~\ref{fig:video-freeway}, the rSVDdpd algorithm outputs a noise free background even if the original video frame is noisy, thus eliminating such undesirable behaviours. The results obtained from all the videos under  these benchmark datasets are presented, for brevity, on the accompanying rSVDdpd webpage https://subroy13.github.io/rsvddpd-home/ (See Section~\ref{sec:website} for details).

\subsection{Application to a Large Scale Camera Tampering Video Surveillance Dataset}\label{sec:real-dataset}

As mentioned in Section~\ref{sec:introduction}, video surveillance under the presence of camera tampering is a very challenging and useful problem. With the problem of camera tampering detection (i.e., classification of tampered frames) in mind,~\cite{mantini2019uhctd} have compiled a comprehensive large-scale dataset called UHCTD (University of Houston Camera Tampering Detection Dataset), which we choose here for the purpose of illustration of performances of these background modelling algorithms under real-life video surveillance data. The dataset contains surveillance videos of over 288 hours ranging across 6 days from two cameras. Three types of camera tampering methods have been synthesized in the dataset; (a) covered, (b) defocused, and, (c) moved (see more details in~\cite{mantini2019uhctd}). These tamperings are done uniformly over the data to capture changing illumination between day and night.

\begin{figure*}[!t]
    \centering
    \subfloat[]{
        \includegraphics[width=0.18\linewidth]{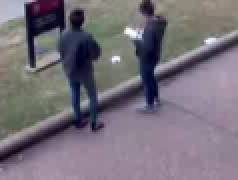}
        \includegraphics[width=0.18\linewidth]{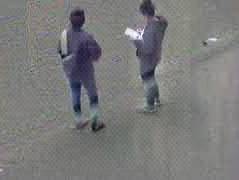}
        \includegraphics[width=0.18\linewidth]{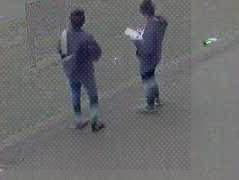}
        \includegraphics[width=0.18\linewidth]{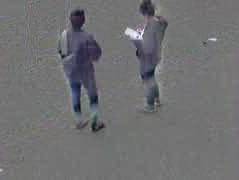}
        \includegraphics[width=0.18\linewidth]{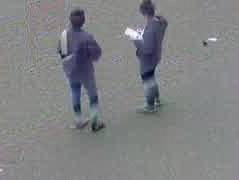}
    }
    \hfill 
    \subfloat[]{
        \includegraphics[width=0.18\linewidth]{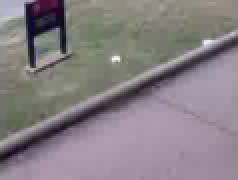}
        \includegraphics[width=0.18\linewidth]{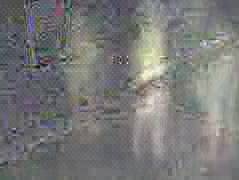}
        \includegraphics[width=0.18\linewidth]{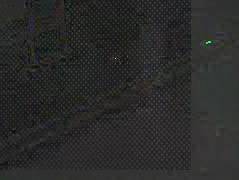}
        \includegraphics[width=0.18\linewidth]{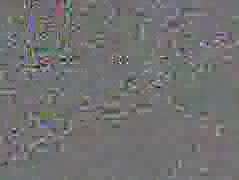}
        \includegraphics[width=0.18\linewidth]{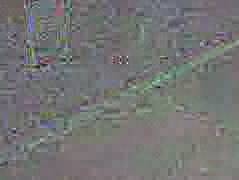}
    }
    \hfill
    \subfloat[]{
        \includegraphics[width=0.18\linewidth]{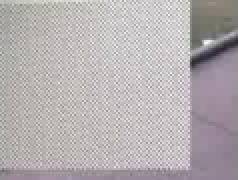}
        \includegraphics[width=0.18\linewidth]{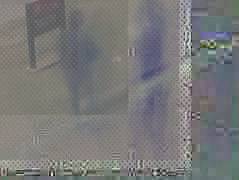}
        \includegraphics[width=0.18\linewidth]{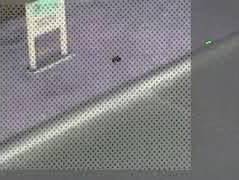}
        \includegraphics[width=0.18\linewidth]{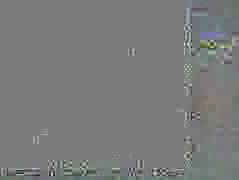}
        \includegraphics[width=0.18\linewidth]{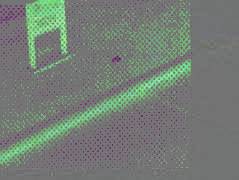}
    }
    \caption{True video frames and the estimated foreground by usual SVD, rSVDdpd, ALM and GoDec algorithms (Left to Right) for frame 5, 50, 75 in UHCTD Day 1 Camera B dataset. Enlarged better quality images are available online at https://subroy13.github.io/rsvddpd-home/}
    \label{fig:video-uhctd}
\end{figure*}

Figure~\ref{fig:video-uhctd} depicts the true frames along with the estimated foregrounds obtained from the usual SVD, robust rSVDdpd, robust PCA via ALM and GoDec algorithms. The estimated backgrounds obtained from other robust PCA methods are visually similar to the output of the ALM algorithm and hence are omitted here. In this surveillance scene from Camera B~\cite{mantini2019uhctd}, a noisy image is synthesized to obstruct the view of the camera. The background estimated from the usual SVD shows a clear indication of the noise even in the frames where camera tampering is not induced, and such an effect is amplified further by the presence of shadow-like artifacts in the estimated background content. Such an effect is also prevalent across the existing techniques. In comparison, the background estimated via the proposed rSVDdpd algorithm removes such artifacts and is less affected by the tampering. This robustness property can also be seen in the estimated foreground content, when the true frame is only the non-tampered background, the proposed rSVDdpd outputs foreground content as a very dark black image without any distinguishing feature as desired. 

\subsection{Further Illustrations}\label{sec:website}

Further illustrations of the performance of the proposed rSVDdpd over the existing comparative algorithms are available in the accompanying rSVDdpd Website https://subroy13.github.io/rsvddpd-home/. The website contains an extended abstract, the necessary \texttt{R} codes, and the detailed description and the images and videos of the estimated foreground of the benchmark datasets and the tampered videos from UHCTD dataset. The proposed method is incorporated in the \texttt{R}-package \texttt{rSVDdpd} and made publicly available via \texttt{CRAN} repository (https://cran.r-project.org/web/packages/rsvddpd/index.html).

\section{Conclusion}\label{sec:conclusion}

In this paper, we had set out to solve the extremely important problem of modelling video surveillance background content robustly, and we trust we have achieved that with the help of a novel algorithm which performs the singular value decomposition robustly. This, we believe, has been adequately demonstrated in our examples. Several existing methods such as GRASTA, MOG and GoDec algorithms which are extensively used for this task, have been shown to have degraded performance in presence of both natural noise and camera tampering~\cite{bouwmans2010statistical,bouwmans2014traditional}. The more recent approaches based on robust PCA perform better than these algorithms, although at the cost of high computational complexity~\cite{bouwmans2014robust}. In comparison, our novel robust singular value decomposition technique based on density power divergence can provide a faster alternative to the robust PCA based methods for the statistical background modelling problems (without compromising robustness). Through a comparative study based on the benchmark dataset, we have demonstrated that the proposed rSVDdpd algorithm is extremely fast and parallelizable, and achieves a performance that is as good as these existing state of the art algorithms. Moreover, in the presence of natural or artificial camera tampering, which is ubiquitous in real-life video surveillance data, the proposed rSVDdpd algorithm yields more reliable background estimates of the non-tampered frames than existing algorithms.

The proposed rSVDdpd algorithm is used on a batch of video frames. However, an online version of rSVDdpd algorithm could be extremely useful to be used in a real-time video surveillance application, where the low rank component can automatically update as soon as new video frames come in. Another direction of future research could be to develop a data-driven method to estimate the rank of the background matrix $\bb{L}$, which will control the amount of slowly moving objects in the background content. Also, as SVD is abundantly used for analyzing high dimensional data, it would be useful to know the range of applications where the proposed rSVDdpd can replace the standard SVD to counter data contamination. We hope to take up these problems in the future.


\bibliographystyle{unsrtnat}
\bibliography{reference-video}  

\end{document}